\documentclass[10pt,conference]{IEEEtran}

\title{\name: Frequency-Aware Component Placement for Superconducting Quantum Computers}
\author{
Junyao Zhang\textsuperscript{1}, 
Hanrui Wang\textsuperscript{2}, 
Qi Ding\textsuperscript{2},
Jiaqi Gu\textsuperscript{3},
Reouven Assouly\textsuperscript{2}\\
William D. Oliver\textsuperscript{2},
Song Han\textsuperscript{2},
Kenneth R. Brown\textsuperscript{1},
Hai "Helen" Li\textsuperscript{1},
Yiran Chen\textsuperscript{1}\\
\textsuperscript{1}Duke University, 
\textsuperscript{2}Massachusetts Institute of Technology,
\textsuperscript{3}Arizona State University,
}

\usepackage{tikz}
\usepackage{physics}
\usepackage{eucal}
\usepackage{xspace}
\usepackage{multirow}
\usepackage{algorithm}
\usepackage{algpseudocode}
\usepackage{enumitem}
\usepackage{caption}
\usepackage{soul}
\usepackage{color}
\usepackage{booktabs} 

\usepackage{cite}
\usepackage{amsmath,amssymb,amsfonts}
\usepackage{graphicx}
\usepackage{textcomp}
\usepackage{xcolor}
\usepackage[hyphens]{url}
\usepackage{fancyhdr}
\usepackage{hyperref}

\usepackage{ulem} 

\hypersetup{
    colorlinks = true,
    linkcolor = blue,
    urlcolor  = blue,
    citecolor = blue,
}

\newcommand{\name}{Qplacer\xspace}
\newcommand{\FH}{Frequency Hotspots\xspace}

\newcommand{\x}{$\times$\xspace}

\newcounter{rlabelno}

\begin{document}

\maketitle

\begin{abstract}

Noisy Intermediate-Scale Quantum (NISQ) computers face a critical limitation in qubit numbers, hindering their progression towards large-scale and fault-tolerant quantum computing. A significant challenge impeding scaling is crosstalk, characterized by unwanted interactions among neighboring components on quantum chips, including qubits, resonators, and substrate.
We motivate a general approach to systematically resolving multifaceted crosstalks in a limited substrate area.
We propose \name{}, a frequency-aware electrostatic-based placement framework tailored for superconducting quantum computers, to alleviate crosstalk by isolating these components in spatial and frequency domains alongside compact substrate design.
\name commences with a frequency assigner that ensures frequency domain isolation for qubits and resonators. It then incorporates a padding strategy and resonator partitioning for layout flexibility. Central to our approach is the conceptualization of quantum components as charged particles, enabling strategic spatial isolation through a `frequency repulsive force' concept.
Our results demonstrate that \name carefully crafts the physical component layout in mitigating various crosstalk impacts while maintaining a compact substrate size. 
On various device topologies and NISQ benchmarks, \name improves fidelity by an average of 36.7\x and reduces spatial violations (susceptible to crosstalk) by an average of 12.76\x, compared to classical placement engines.
Regarding area optimization, compared to manual designs, \name can reduce the required layout area by 2.14\x on average.

\end{abstract}

\section{Introduction}\label{sec:intro}
In the ever-evolving realm of quantum computing, quantum computers (QCs) in Noisy Intermediate-Scale Quantum Computing (NISQ) \cite{NISQ} have captivated people's attention, offering immense potential of computation to unravel intricate problems in chemistry \cite{VQE_chemistry, chemistry}, biology \cite{biology}, algorithms \cite{grover, shor}, and machine learning \cite{QML, quantumnas}. Superconducting QCs \cite{SC, SC_2}, particularly the utilizing fixed-frequency transmon qubits anchored by Josephson junctions (JJs), stands out as a leading option for scalable quantum computing. These qubits contribute to relative rapid gate speeds \cite{gate_time1, gate_time2, quantum_progress}, fairly extended coherence duration \cite{coherence_time} and gate fidelities nearing fault-tolerance thresholds \cite{cr_gate}. These merits have led to significant advancements in the field, as evidenced by the current generation of superconducting QCs, which boasts over 1000 physical qubits \cite{ibm_1000}. However, this scale of QCs is still insufficient for tackling intricate real-world problems more efficiently and outpacing classical computers\cite{scale}.

In the near term, the critical barrier encountered in scaling superconducting QCs is crosstalk, a prevalent issue in many quantum architectures \cite{ionq, SC, google}. 
Crosstalk arises from unwanted interactions between components on a quantum chip, often triggered when elements with resonating frequencies are either connected or positioned in close proximity \cite{crosstalk, para_g, para_g_2}. In superconducting QCs, this leads to compromised computational fidelity, with inter-qubit crosstalk being a focal concern. However, crosstalk impact extends beyond qubit interactions to resonators, which are integral for entangling qubits, facilitating qubit interactions, and reading qubit states. Crosstalk between resonators can inadvertently affect gate operations, undermining qubit fidelity \cite{res_crosstalk, physical_process_2, device_crosstalk, substrate_limit_2}.

Additionally, interactions between quantum components, such as qubits and resonators, and their substrate present another layer of complexity.
The strong electromagnetic field coupling intrinsic to superconducting qubits becomes problematic with increased substrate size. This enlargement triggers spurious electromagnetic modes that lower the substrate frequency, leading to substrate-qubit crosstalk, inter-qubit crosstalk, and diminishing coherence times \cite{substrate_limit, spurious, quantum_progress, dist_modeling}. 
Thus, the challenge in scaling QCs lies in expanding the qubit array while maintaining frequency or spatial isolation to reduce inter-component (qubits and resonators) crosstalk and keeping the substrate size compact to prevent spurious modes \cite{crosstalk_twoQ, crosstalk_ding, LOM, substrate_limit, substrate_limit_2}.

\begin{figure}[t]
  \centering
  \includegraphics[width=0.95\linewidth]{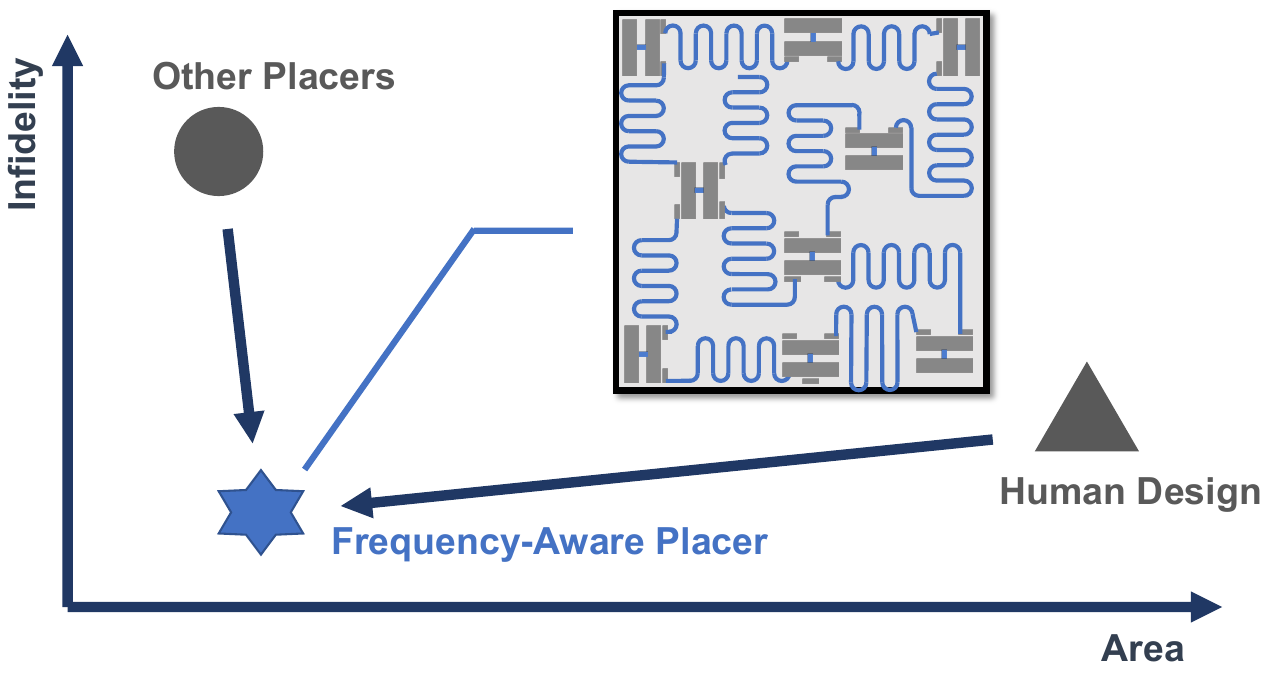}
  \vspace{-10pt}
  \caption{
  System infidelity due to crosstalk impacts versus the area required for accommodating an equal number of qubits and other quantum components using different placement strategies. \name{} is designed to optimize layout area while maintaining low infidelity.
  }
  \label{fig:teaser}
  \vspace{-10pt}
\end{figure}

Substantial strides have been made in countering crosstalk, with the primary focus on averting unintended resonance between qubits. Current strategies predominantly include the use of compilers and schedulers in systems with fixed couplings \cite{ibmq, fix_xtalk_miti}, as well as the integration of tunable components in more adaptive architectures \cite{crosstalk_ding, gate_time1, tunable_qubit, tunable_qubit_2, google}, 
These methods primarily focus on temporal or frequency domain isolation to avoid crosstalk \cite{crosstalk, crosstalk_ding, ibm_crosstalk}.
However, this concentration on specific forms of crosstalk is often insufficient in the system design of processor and has left other critical aspects underexplored. As quantum systems evolve in complexity and scale, a comprehensive understanding and effective mitigation of diverse forms of crosstalk become imperative. Addressing these challenges is essential for enhancing the QC robustness.

In this study, we systematically investigate and address the multifaceted challenges associated with spatial and frequency constraints in inter-qubit and resonator crosstalk, as well as keeping a compact substrate size to suppress substrate crosstalk. Recognizing the complexity of these issues, our solution pivots on a meticulously
designed physical layout that not only scales up QCs but also preserves system fidelity amid these multifaceted interactions. This initiative marks a pivotal advancement in quantum chip design, harmonizing the delicate balance between component isolation and integration in sophisticated quantum systems.

To achieve these, we introduce a \textbf{frequency-aware electro-static-based analytical placement framework}, \name{} (analytical placement refers to the process of determining the optimal physical positions of various components on a substrate; we use placement for short). 
\name{} begins with a frequency assigner that allocates distinct frequencies to qubits and resonators, ensuring frequency domain isolation for all interconnected quantum components.
We then implement a padding strategy for movable quantum components to setup minimum spacing, and partitioning resonators space into smaller segments for greater flexibility in layout design. 
Next, we draw a parallel between the quantum device components (such as qubits and resonators), which possess specific frequency properties, and charged particles to strategically position these components. 
This analogy allows the application of what we term `\textbf{frequency repulsive force}', a novel concept that operates exclusively between components sharing similar frequencies. Leveraging this principle enables the strategic positioning of these components, ensuring they are sufficiently distanced from each other spatially.
Finally, our framework includes a legalizer step, which integrates the segmented resonators to ensure their integrity.
This comprehensive approach not only diminishes crosstalk but also maximizes the scalability of the quantum chip. By optimizing the use of available space, \name{} results in a compact layout. Fig.\ref{fig:teaser} illustrates a comparison of with classical methods and human-designed layouts, highlighting the advancements of \name{}.

Our electrostatic-based placement approach, inspired by classical placement framework \cite{eplace, eplace_ms, replace, dreamplace3}, distinguishes itself with three major differences: 
(1) We integrate frequency penalties into our cost function to enhance the versatility of models, making it keenly responsive to crosstalk impacts and thereby bolstering the system robustness;
(2) To address the unique spatial demands of quantum devices, specifically resonators, we introduce padding and partitioning strategies tailored for quantum-specific needs;
(3) We also introduce a specialized legalization process, adept at managing the diverse shapes of quantum components and segmented resonators, a notable departure from the uniformity commonly observed in classical circuit designs.

\noindent The contributions of this work are summarized as follows:
\setlist{nolistsep}
\begin{itemize}[leftmargin=*]
    \item To our knowledge, this research is the first to comprehensively address the impacts of resonator and substrate crosstalk, along with spatial constraints in quantum components, for the assessment of quantum system fidelity and scalability.
    \item We propose \name, a framework designed to 
    meticulously craft the physical layout of superconducting QCs effectively mitigating various crosstalk impacts while optimizing for a compact substrate size.
    \item To achieve this, \name designs \textbf{Frequency-Aware Electrostatic-based Placement Engine}, which envisions quantum components as particles, implements a repulsion mechanism for spatially isolating components with similar frequencies.
    \item \name proposes \textit{padding technique}, \textit{resonator partitioning}, and an \textit{integration legalizer}, facilitating the placement process and addressing the unique challenges of quantum component layout.
    \item With techniques above, \name enhances fidelity by an average of 36.7\x and reduces spatial violations—areas susceptible to crosstalk—by an average of 12.76\x compared to classical placer. For area optimization, \name reduces the layout area by an average of 2.14\x over manual designs.
\end{itemize}

\section{Background}
\subsection{Transmon Qubits}\label{sec:qubit}
Transmon qubits are a prominent type of superconducting qubits, integral to various widely used quantum computer architectures \cite{quantum_progress, SC_2, ibm_127_1, ibm_127_2, google, quantware}. In superconducting QCs, entanglement between transmon qubits is employed using physical coupling mechanisms such as capacitors \cite{crosstalk_twoQ, intro_cQED}, resonators (linear couplers) \cite{cQED, intro_cQED}, and tunable couplers \cite{tunable_coupler, google}. This work primarily focuses on fixed-frequency transmon architectures coupled by resonators \cite{SC, SC_1}, as shown in Fig.\ref{fig:gate}. The resonator is a quantum harmonic oscillator composed by a linear inductor and capacitor.

\begin{figure}[t]
  \centering
  \includegraphics[width=\linewidth]{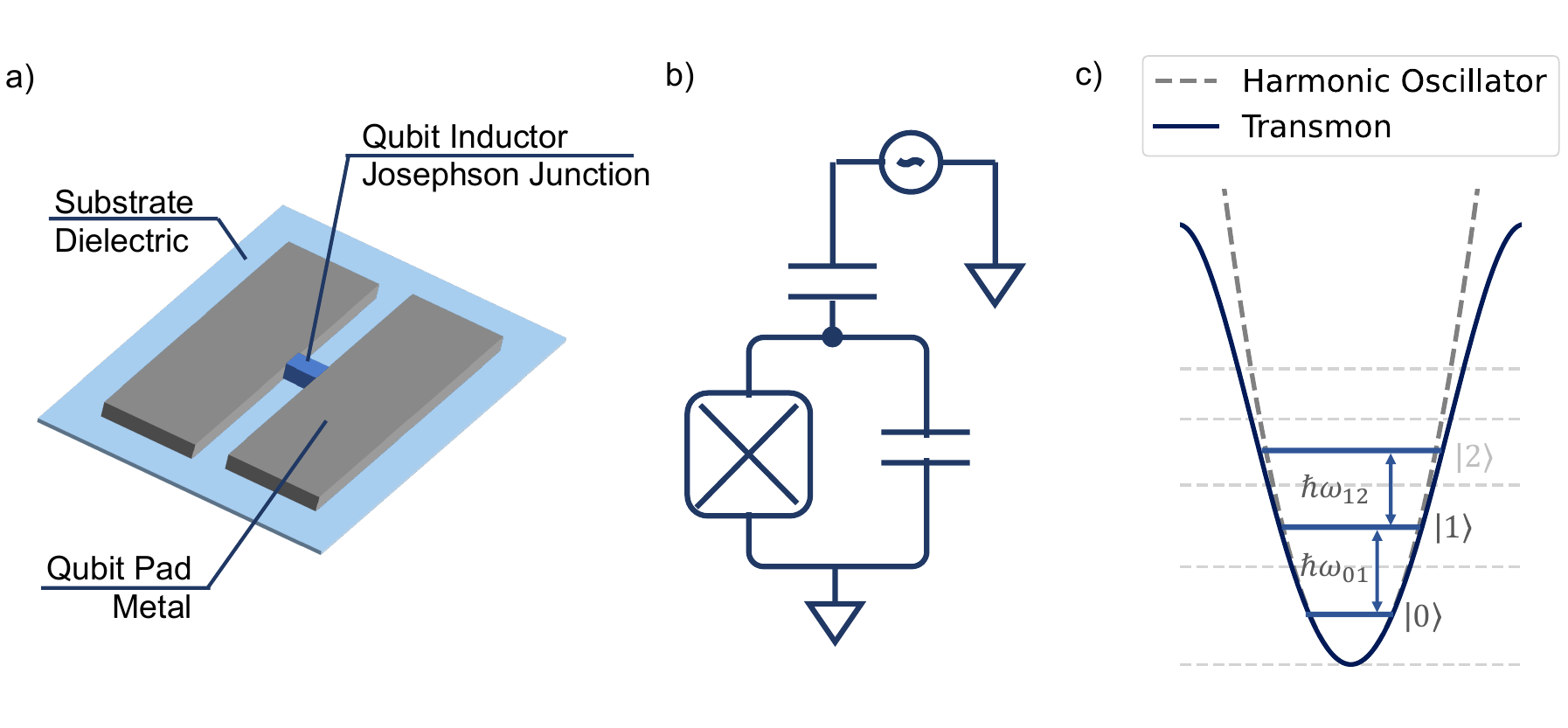}
  \caption{
  \textbf{a):} Physical layout of a transmon qubit. 
  \textbf{b):} Circuit diagram of a fixed-frequency transmon qubit featuring a capacitor, Josephson junction, and microwave control line. 
  \textbf{c):} Energy levels of transmon, Josephson junction transforms energy potential from quadratic (dashed dark gray) to sinusoidal (solid blue), leading to distinct energy levels $|0\rangle$ and $|1\rangle$ for computational use, with energy separation $\hbar \omega_{01}$.
  }
  \label{fig:transmon}
  \vspace{-5pt}
\end{figure}

Fig.\ref{fig:transmon}-a illustrates the physical layout of a transmon qubit. The substrate, usually a dielectric material like silicon, supports two metallic pads connected via a non-linear inductor (Josephson junctions). These components are lithographically printed on the substrate, forming the transmon qubit, a quantum anharmonic oscillator. To facilitate inter-qubit couplings, additional smaller metal pads can be added for connections. Circuit schematic is illustrated in Fig.\ref{fig:transmon}-b.

Transmon qubit operates as a multi-level quantum system designed to exhibit an atom-like energy spectrum, as depicted in Fig.\ref{fig:transmon}-c. The two lowest energy states, representing binary 0 and 1, are the ground state ($|0\rangle \equiv [1;0]^T$) and the first excited state ($|1\rangle \equiv [0;1]^T$). Unlike classical bits, a qubit can exist in a superposition of these states, formulated as $|\phi\rangle = \alpha|0\rangle + \beta|1\rangle = [\alpha;\beta]^T$, where $\alpha$ and $\beta$ are complex coefficients satisfying the normalization $|\alpha|^2 + |\beta|^2 = 1$.
The qubit frequency, $\omega_q$, is defined by the energy gap between the ground state and the first excited state, expressed as $E_{01} =\hbar \omega_q = \hbar \omega_{01}$ (where $\hbar$ is Planck's constant).

\subsection{Gate Operations}\label{sec:gate}

Quantum computation relies on qubit gates, operations that transition a qubit between quantum states through unitary transformations, represented mathematically as $|\phi\rangle\rightarrow U|\phi\rangle$, where $U$ is a unitary matrix. This section focuses on gate operations within the fixed-frequency transmon architecture.

\textbf{Single Qubit Gates:} These are implemented by modulating qubits with time-dependent microwave voltage signals. The process, depicted in Fig.\ref{fig:transmon}-b, involves a microwave drive line connected to the qubit via a capacitor. Altering the microwave pulses frequency, phase, and amplitude allows for various single qubit gates execution \cite{crosstalk_twoQ, cQED}.

\textbf{Two-qubit Gates:} Two-qubit gates are crucial in quantum computation as they provide entangling operations, allowing transformations of one qubit to be conditional on the state of another qubit \cite{crosstalk_twoQ, cQED, SC}. 
In fixed-frequency transmon architectures, microwave drives activate qubit interactions through all-microwave-based two-qubit gates. Advantages of this approach include longer gate lifetimes, simplified control as single-qubit gates, and minimized crosstalk \cite{detune_flux, detune_flux_2}.
Resonator induced phase gate (RIP) is the leading technique for implementing these microwave-based gates, compared with other candidates \cite{cr_gate, map_gate}. It does not only retains all the benefits of all-microwave-based two-qubit gates but also is adaptable to a broad range of qubit spectral profiles \cite{rip_gate, rip_gate_2, rip_gate_3, rip_gate_4}. 

\begin{figure}[t]
  \centering
  \includegraphics[width=0.9\linewidth]{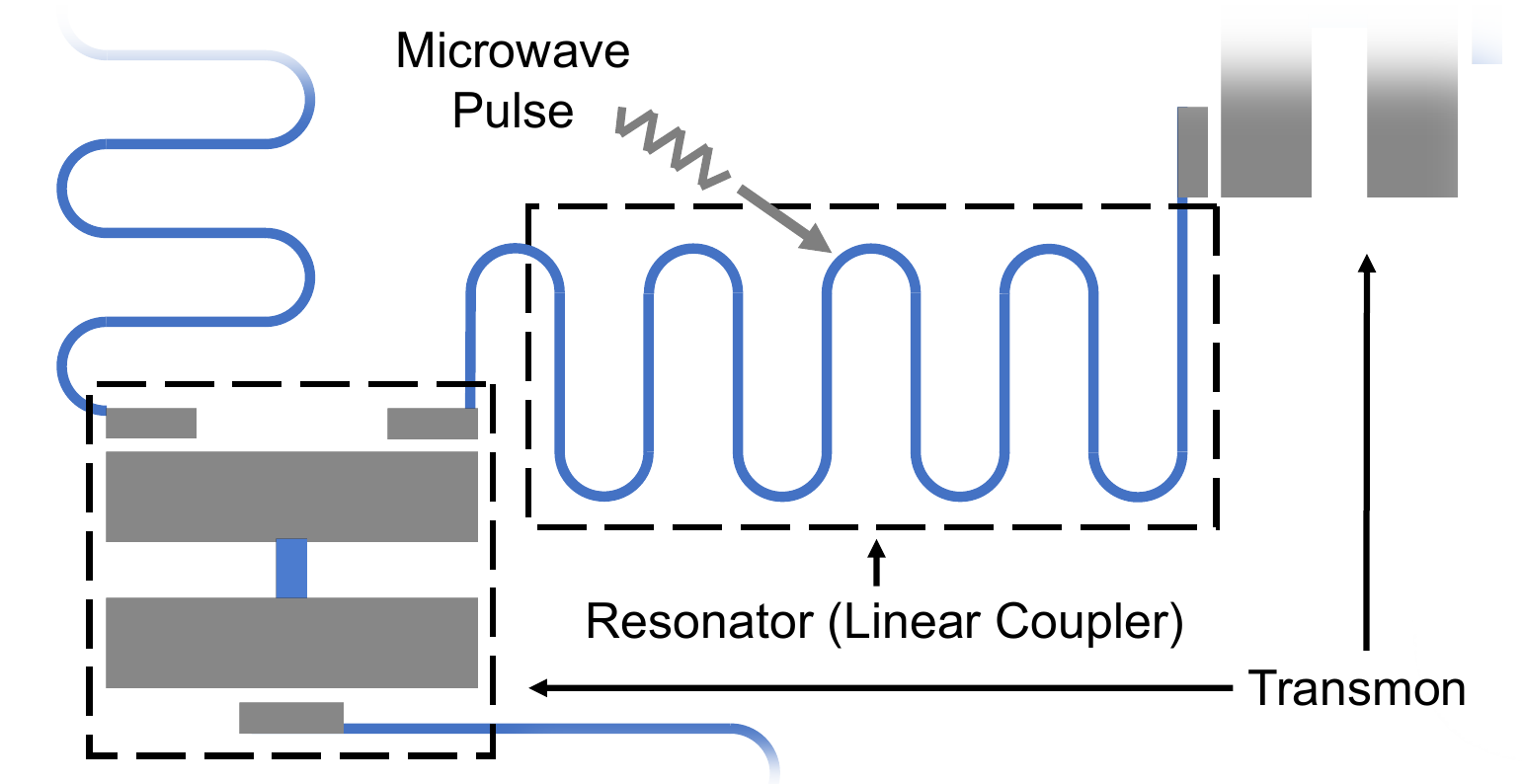}
  \caption{
  Circuit diagram of two coupled transmon qubits via a resonator; Two-qubit gates (\textit{CZ}, controlled Phase gate) are implemented by applying/removing an off-resonant pulse to the resonator.
  }
  \label{fig:gate}
  \vspace{-5pt}
\end{figure}

The control mechanism of a RIP gate is illustrated in Fig.\ref{fig:gate}. 
It operates by coupling two fixed-frequency qubits to a detuned resonator. The gate operation involves applying an off-resonant pulse to the resonator, inducing a phase shift in the qubits without altering the resonator. This mechanism enables the implementation of a Controlled-PHASE (\textit{CZ}) gate. Mathematically, two-qubit RIP gate operation is described as:
\begin{equation}
    U = exp [-i \dot \theta \sigma_{z} \otimes \sigma_{z} t]
\end{equation}
Here, $\sigma_{z} \otimes \sigma_{z}$ represents the joint operation on both qubits, and $\dot \theta$, the coupling rate, scales according to:
\begin{equation}\label{eq:coupling_rate}
    \dot{\theta} \propto \underbrace{\left(\frac{|\Omega V_d|}{2\Delta_{cd}}\right)^2}_{\bar{n}} \frac{\chi}{\Delta_{cd}}
\end{equation}
where $\bar{n}$ is the average photon count in the resonator, $\chi$ the dispersive shift $\chi = g^2/|\omega_r-\omega_q|$, $\omega_r$ and $\omega_q$ represents the frequency of resonator and frequency of qubit, respectively, and $\Delta_{cd}$ the detuning between the drive and the resonator. The Controlled-Z (\textit{CZ}) gate is accomplished by adjusting the product of $\dot{\theta}t = \pi/4$. Enhancing $\dot{\theta}$ is an effective approach to reducing the gate time $t$ and thus achieving faster gate operations.

\section{Crosstalk Challenges}\label{sec:problem}

This section illustrates three primary sources of crosstalk in transmon QCs that hinder the development of larger-scale systems as mentioned in Section \ref{sec:intro}. These include crosstalk between qubits, interference within resonators, and substrate spurious mode impact. The intensity of crosstalk noise is notably influenced by the strength of interactions between qubits \cite{cQED, crosstalk_twoQ, crosstalk_ding}. To effectively quantify the impact of crosstalk, we focus on the coupling strength as a representative measure and systematically analyze it using the Jaynes–Cummings Hamiltonian \cite{jc_hamil, jc_hamil_2, jc_hamil_3}. 

\subsection{Inter-Qubit Crosstalk}\label{sec:q_x}

Inter-qubit crosstalk presents a significant challenge in the scaling of superconducting quantum architectures. 
This issue is primarily rooted in the unintended interactions between qubits, especially when they are either directly coupled via capacitor or in spatial close proximity. 
For instance, when two qubits are near resonance, their dynamics can be represented by a Hamiltonian model that includes individual system Hamiltonians $H_1$ and $H_2$, and an interaction Hamiltonian $H_{int}$ as shown in Eq.(\ref{generic_H}): 
\begin{equation}\label{generic_H}
    H = H_1 + H_2 + H_{int}
\end{equation}
Particularly, when qubits are at or near resonance (\( \omega_1 \approx \omega_2 \)), their interaction is described by Hamiltonian $H_{\text{res}}$:
\begin{equation}
    H_{\text{res}} = \omega_1 \sigma_z^1 + \omega_2 \sigma_z^2 + g (\sigma_+^1 \sigma_-^2 + \sigma_-^1 \sigma_+^2) 
\end{equation}
Where $\omega_i$ represents the frequencies of the $i$-th qubit. The terms $\sigma_z^i$, $\sigma_+^i$, and $\sigma_-^i$ denote the Pauli Z matrices and the raising and lowering operators for the $i$-th qubit, while $g$ is the coupling strength.

\begin{figure}[t!]
    \centering
    \includegraphics[width=0.9\linewidth]{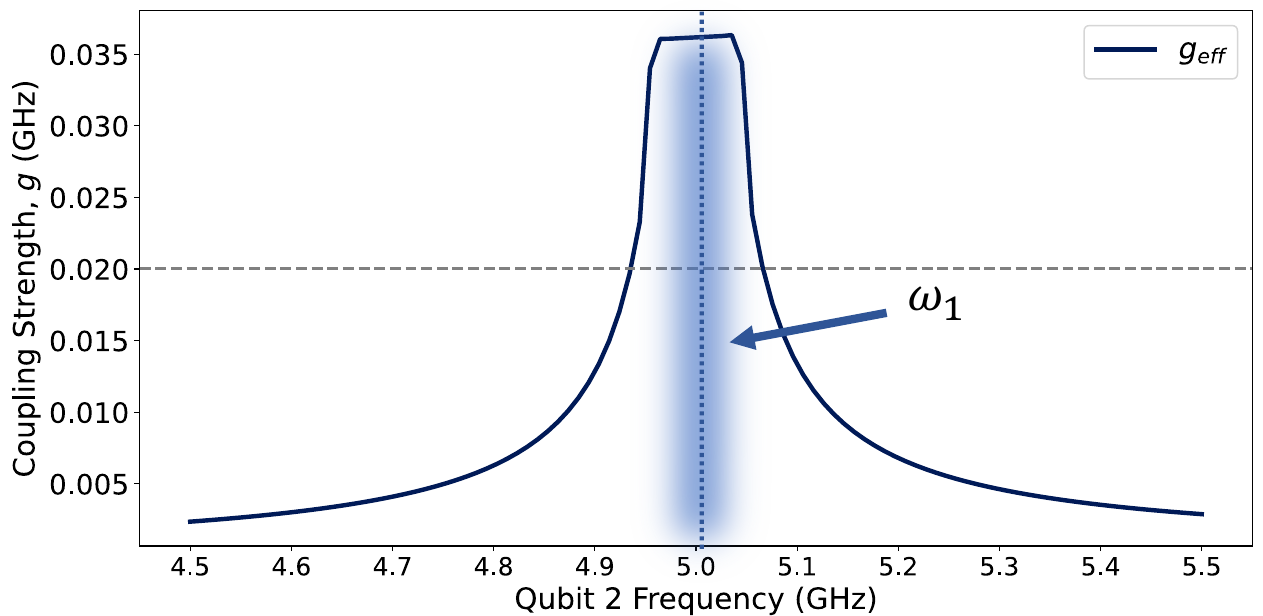}
    \caption{
    Coupling strength between two directly connected transmon qubits via a capacitor. $\omega_n$ for qubit $n$. $\omega_1$ is held constant while $\omega_2$ is varied. The peak coupling strength occurs when the two transmons are resonant ($\omega_1 = \omega_2$), depicted in blue shadow. As $\omega_2$ diverges from $\omega_1$, the residual coupling gradually diminishes. Coupling strength $g$ is typically around $20 \sim 30$MHz (gray dash line).
    }
    \label{fig:g_q_q}
    \vspace{-5pt}
\end{figure}

For significantly detuned scenarios ($\Delta = |\omega_1 - \omega_2| \gg g$), Hamiltonian $H_{\text{off-res}}$ applies \cite{cQED, crosstalk_twoQ}:

\begin{equation} \label{eq:H_qq}
    H_{\text{off-res}} \approx \omega_1 \sigma_z^1 + \omega_2 \sigma_z^2 + g_{\text{eff}}\sigma_z^1 \sigma_z^2
\end{equation}
showing a reduced effective coupling strength $g_{\text{eff}} = g^2/\Delta$. Fig.\ref{fig:g_q_q} illustrates the above interactions indicating that effective coupling can be modulated by adjusting qubit frequencies.

Existing research predominantly focuses on mitigating crosstalk by fine-tuning the frequencies of directly connected qubits \cite{crosstalk_ding, crosstalk_twoQ, ibm_crosstalk}, thus reducing residual coupling and preventing the system from entering chaotic states \cite{chaos, para_g}. However, such approaches do not fully address crosstalk in finite-sized circuits where physical distance between qubits introduces parasitic capacitive couplings \cite{para_g, para_g_2, para_g_3}, a factor often overlooked in previous studies.

\begin{figure}[t!]
    \centering
    \includegraphics[width=\linewidth]{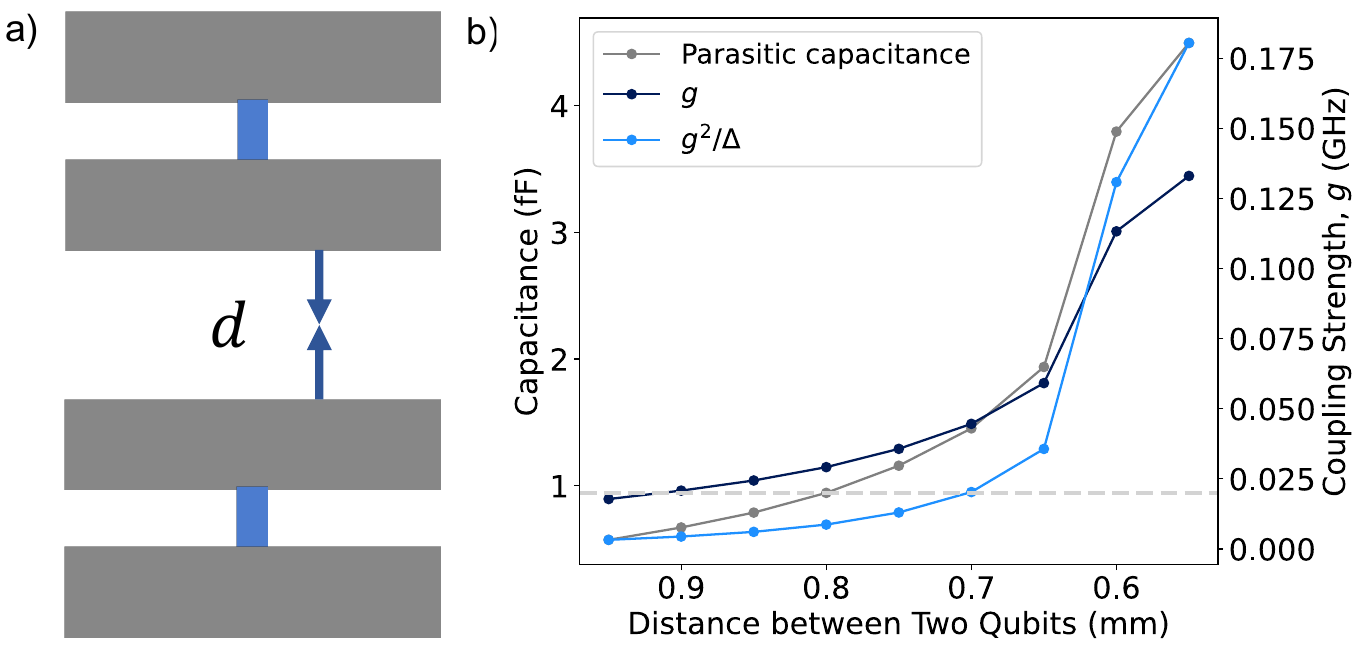}
    \caption{
    \textbf{a):} Separation distance $d$ between transmon qubits. 
    \textbf{b):} Variation in coupling strength $g$, effective coupling strength $g_{\text{eff}}$ and parasitic capacitance $C_p$ with the distance $d$ between two transmon qubits, indicating increased capacitance and coupling strength as $d$ decreases.
    }
    \label{fig:g_q_dist}
\end{figure}

To analyze these unwanted parasitic capacitive couplings, we consider scenarios with qubits in spatial close proximity with distance $d$, as depicted in Fig.\ref{fig:g_q_dist}-a. The Hamiltonian of such a system, analogous to Eq.(\ref{eq:H_qq}), now includes the parasitic coupling strength $g$ which is defined as \cite{crosstalk_twoQ}:
\begin{equation}\label{eq:g_qq}
    g = \frac{1}{2}\sqrt{\omega_1 \omega_2} \frac{C_p}{\sqrt{C_1+C_p}\sqrt{C_2+C_p}}
\end{equation}
Where $C_p$ is the parasitic capacitance between two qubits and $C_i$ is the capacitance of qubit $i$. Eq.(\ref{eq:g_qq}) shows parasitic coupling strength $g$ depends on parasitic capacitance $C_p$, which is inversely related to the distance between qubits.
Fig.\ref{fig:g_q_dist}-b shows the simulated $C_p$ value from Qiskit Metal \cite{Qiskit_Metal, Qiskit, LOM}, which demonstrates an increase in parasitic capacitance as qubits are positioned closer, leading to a rise in both parasitic coupling strength $g$ for two resonated qubits and effective coupling $g^2/\Delta$ for two detuned qubits. Therefore, effective mitigation of inter-qubit crosstalk requires assigning distinct frequencies to each pair of connected qubits and maintaining adequate physical separation, particularly for qubits with similar frequencies.

\subsection{Resonator Crosstalk}\label{sec:res_x}

In Section \ref{sec:q_x}, we discuss inter-qubit crosstalk, highlighting the need for isolation in both spatial and frequency domains. 
However, the available frequency spectrum for qubits is typically constrained to around 5GHz \cite{ibmq, quantum_progress}, limited by peripheral control device costs and thermal noise vulnerability \cite{env_setup, rip_gate, crosstalk_twoQ}. Limited frequency spectrum leads to ``frequency crowding" in larger systems or programs employing parallelism, complicating the frequency assignment. Furthermore, architectures with direct capacitive coupling require significant tuning of qubit frequencies or the use of tunable couplers. These adjustments can introduce additional dephasing and risk accidental resonance with other qubits or environments, potentially resulting in crosstalk \cite{cQED, intro_cQED}.

In addressing the constraints of direct capacitor coupling and relief the frequency crowding, using resonators as quantum buses emerges as an effective alternative for mediating inter-qubit crosstalk, offering selective coupling advantages \cite{cQED, cQED_2}. The Hamiltonian of qubit-resonator interaction is:
\begin{equation}
    H = \frac{\omega_q}{2} \sigma_z + \omega_r a^\dagger a + g (\sigma_+a + \sigma_-a^{\dagger}) 
\end{equation}
where $a^\dagger$ and $a$ are the creation and annihilation operators for resonator, $\omega_q$ and $\omega_r$ are qubit and resonator frequency, respectively. Resonators provide distinct operational regimes. In the vacuum Rabi oscillation regime, strong energy exchange occurs when qubit and resonator frequencies are close. Conversely, in the dispersive regime where $\Delta = |\omega_q - \omega_r| \gg g$, the Hamiltonian approximates to: 
\begin{equation}
    H \approx \frac{\Tilde{\omega_q}}{2} \sigma_z + \omega_r a^\dagger a + \chi \sigma_z a^\dagger a
\end{equation}
$\chi=g^2/\Delta$ denotes the effective qubit-resonator coupling strength. In this regime, energy transfer is negligible \cite{intro_cQED, cQED}.

\begin{figure}[t]
    \centering
    \includegraphics[width=\linewidth]{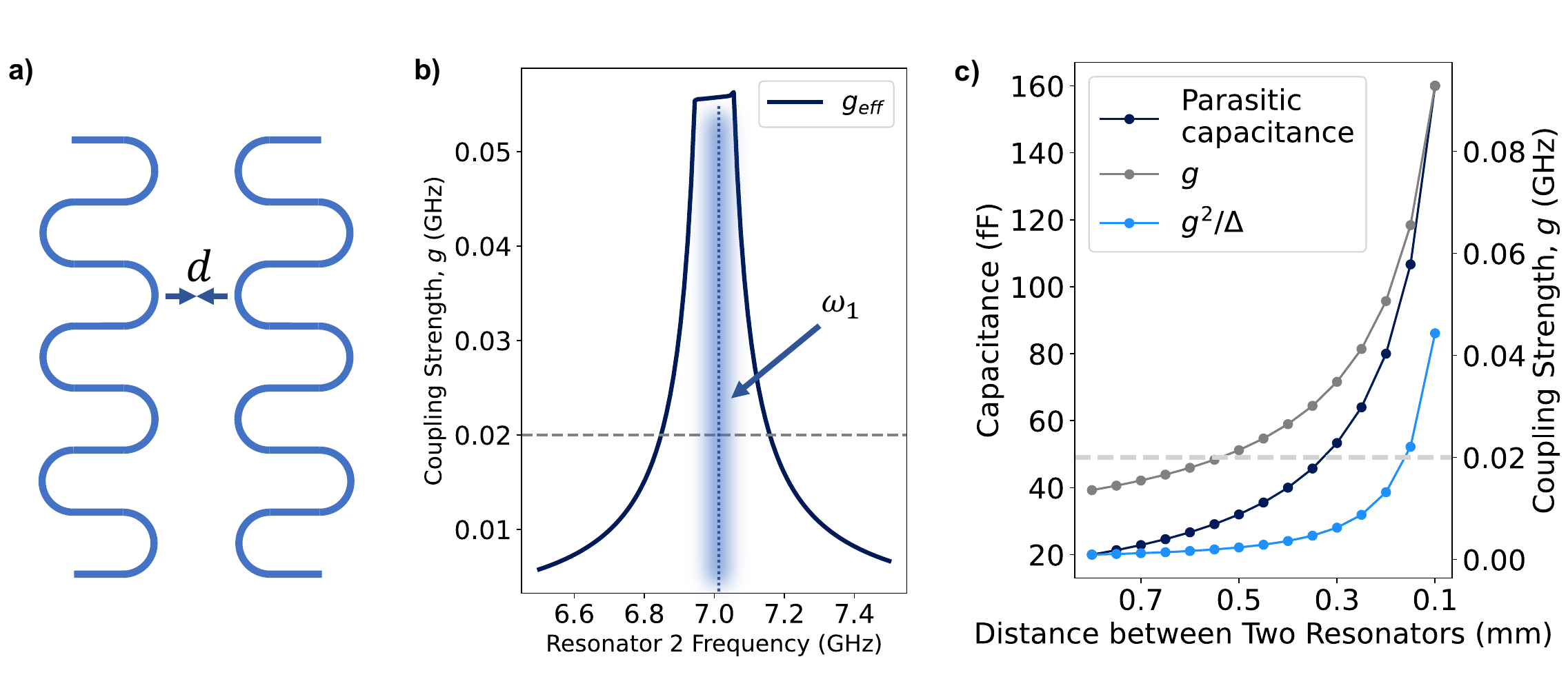}
    \caption{
    \textbf{a):} Diagram illustrating the distance between resonators
    \textbf{b):} Maximum coupling strength at resonator resonance ($\omega_{r1} = \omega_{r2}$).
    \textbf{c):} Coupling strength and parasitic capacitance versus distance $d$ between two resonators, as shown in \textbf{(a)}. Decreased $d$ leads to increased coupling strength.
    }
    \label{fig:g_res_res}
    \vspace{-10pt}
\end{figure}

Although resonator is beneficial in mitigating inter-qubit crosstalk, it presents other challenges. As quantum oscillators, resonators can inadvertently couple with others at or near resonance frequencies, leading to potential crosstalk \cite{res_crosstalk, rabi_split}. The coupling strength escalates from $g^2/\Delta$ to $g$ with narrowing frequency detuning. Additionally, close proximity between resonators also induces parasitic capacitive coupling, described by $g \propto {C_p}/{\sqrt{C_r^1 C_r^2}}$, where $C_p$ is the parasitic coupling capacitance and $C_r^i$ is the $i$-th resonator capacitance \cite{resonator_calcu}. Fig.\ref{fig:g_res_res} illustrates these effects, showing both the frequency-dependent coupling strength and the rise in parasitic coupling as resonators come closer \cite{resonator_calcu, res_crosstalk, resonator_crosstalk}. These unintended energy exchanges between resonators pose more severe challenges than inter-qubit crosstalk. They not only affect qubit fidelity and gate operations but also complicate error correction efforts by violating principles of error locality and independence \cite{surface_code, crosstalk_qec}.

Moreover, resonators consume substantial substrate area due to its considerable wirelength as shown in Fig.\ref{fig:gate}, which imposes additional spatial constraints in scaling up QCs. Therefore, addressing the complexities of resonator crosstalk necessitate meticulous frequency assigning and strategic layout designs in balancing the need for more qubits against the challenges in resonator placement.

\subsection{Substrate Spurious Electromagnetic Mode}\label{sec:sub_x}
Substrate limitations is also a critical concern in scaling up superconducting QCs. Increasing the substrate size to accommodate more qubits is not a straightforward solution in NISQ systems, particularly due to the emergence of spurious electromagnetic modes (box modes) that arise with increased substrate size \cite{substrate_limit, spurious}.
These modes enforce a frequency constraint on the components of the chip, not allowing them to exceed the first eigenmode (TM110) of the substrate \cite{substrate_limit_2, substrate_limit}. For instance, TM110 drops from 12.41 GHz to 6.20 GHz when increasing from a 5$\times$5 $mm^2$ to a 10$\times$10 $mm^2$ substrate size, limiting the available frequency spectrum for qubits and resonators \cite{substrate_limit_2, substrate_limit}. While efforts have been made to mitigate these modes \cite{spurious, dist_modeling, cpw_spur}, the physical size of superconducting chips is typically limited to about 10$\times$10 $mm^2$ \cite{substrate_limit, substrate_limit_2}. Additionally, substrate area directly influences manufacturing costs and fridge requirements\mbox{\cite{quantum_progress, quantum_progress_2}}. Therefore, enhancing the utilization of available substrate space through innovative analytical placement techniques becomes a viable approach to scale up quantum computers within these physical constraints.

\begin{figure*}[ht!]
  \centering
  \includegraphics[width=\linewidth]{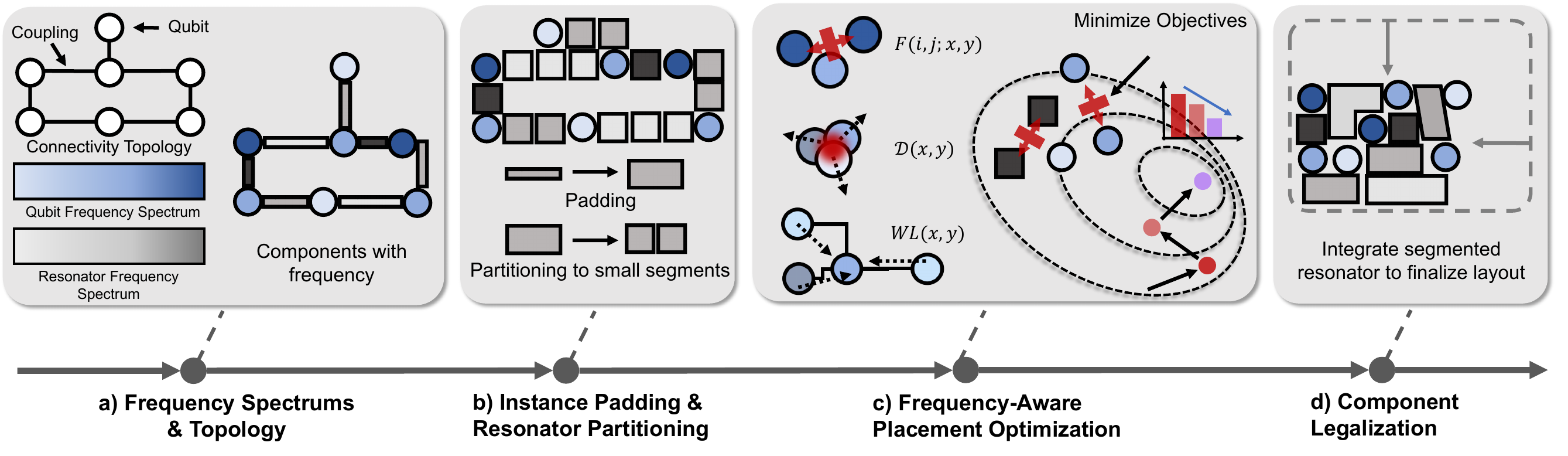}
  \caption{
      Overview of \name{}.
      \textbf{a}: Target connectivity topology and frequency allocation for components (qubits and resonators), based on available frequency spectrums.
      \textbf{b):} Padded qubits and resonators, with resonators segmented for layout flexibility
      \textbf{c):} Frequency-aware placement engine optimizing components' position for crosstalk mitigation, area minimization, and density balance using \textbf{frequency repulsive forces}.
      \textbf{d):} Integration legalizer finalizes layout, ensuring resonator coherence.
    }
  \label{fig:overview}
  \vspace{-10pt}
\end{figure*}

\section{Frequency-Aware Analytical Placement}
\subsection{Overview}

The proposed placement framework, as illustrated in Fig.\ref{fig:overview}, outlines a systematic approach for optimizing the layout of quantum processors. It requires two primary inputs: a target connectivity topology and the allocation of frequencies to components, determined by the available frequency spectrums.
The initial stage involves quantum-specific placement preprocessing, where movable components are equipped with padding and resonators space are divided into smaller segments for better management.
Subsequently, the framework conceptualizes quantum device components, each with unique frequency characteristics, as charged particles. 
The placement engine then optimizes the locations of these particles (quantum components) and disperses them across the substrate by balancing the frequency repulsive forces of the components. The process concludes with the integration of the segmented resonator, establishing the final layout configuration. 

\begin{figure}[t]   
  \centering
  \includegraphics[width=\linewidth]{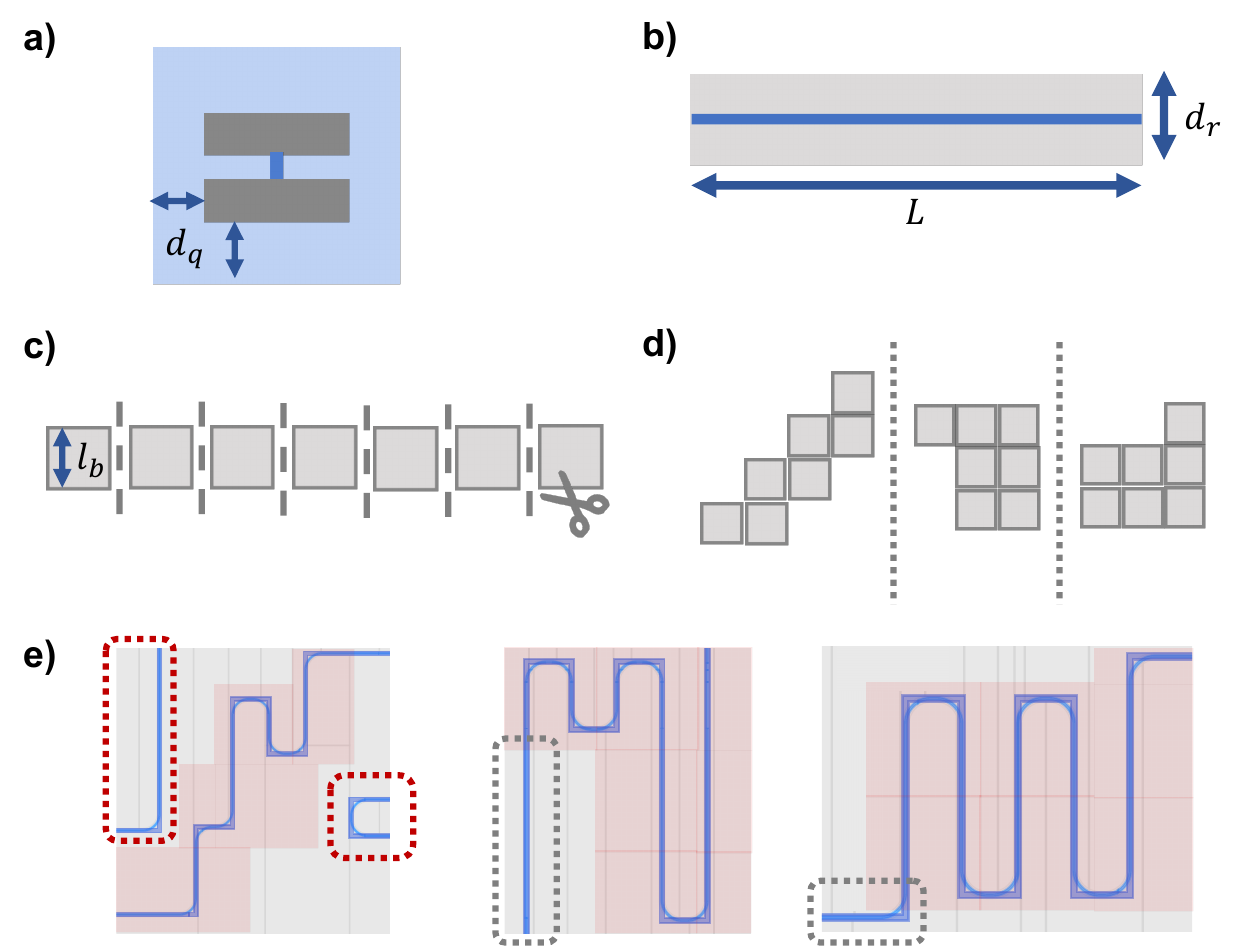}
  \vspace{-10pt}
  \caption{
   \textbf{a):} Qubit padded with distance $d_q$.
  \textbf{b):} Padded resonator with wirelength $L$ and padding distance $d_r$.
  \textbf{c):} Reshaped resonator space 
  from \textbf{(a)}, transformed into a compact rectangle, partitioned into segments of size $l_b$, retaining frequency consistency as indicated by color.
  \textbf{d):} Flexible placement of resonator space 
  segments to accommodate diverse layout designs.
  \textbf{e):} Generated resonator routing using Qiskit Metal \cite{Qiskit_Metal} based on the segmented placement (red shadow) in \textbf{(d)}. red dotted box: obstacle components, gray dotted box: connections between meander resonators and qubit pins.
  }
  \label{fig:pad_parti}
  \vspace{-15pt}
\end{figure}


\subsection{Quantum Specific Placement Preprocessing}

\subsubsection{Quantum Components Padding}
Padding is an essential technique for establishing minimum spacing between quantum components to mitigate crosstalk in quantum layout. As outlined in Section \ref{sec:problem}, close proximity of quantum components can lead to parasitic capacitive coupling, consequently introducing unwanted crosstalk, as demonstrated in Fig.\ref{fig:g_q_dist} and \ref{fig:g_res_res}-c. Both the direct coupling strength $g$ and the effective coupling strength $g^2/\Delta$ surge to undesirable levels when separation distance $d$ is small, leading to energy leakage even among detuned components \cite{para_g, res_crosstalk}. 

In placement optimization, placement engine might inadvertently position components in close adjacency, potentially causing crosstalk issues. To counteract this, padding intentionally adding extra space around each quantum component. This method establishes predefined minimum distances, denoted as $d_q$ for qubits and $d_r$ for resonators, ensuring adequate spatial separation between each component and its neighbors. It is visually represented in Fig.\ref{fig:pad_parti}-a and \ref{fig:pad_parti}-b.

\subsubsection{Resonator Partitioning} \label{sec:partition}

Resonator partitioning is another innovative technique to enhance the flexibility and scalability of QCs, in addressing the challenge of resonator area overhead, as highlighted in Section \ref{sec:res_x}. This technique involves dividing the required space allocated for each resonator into smaller modular segments. 
The process starts by reshaping the allocated resonator  space into a compact rectangle of equivalent area, followed by defining a basic wire block size, $l_b$. Each reshaped resonator space is subsequently segmented based on this predetermined block size, as illustrated in Fig.\ref{fig:pad_parti}-c. This strategy ensures the preservation of the resonators' fundamental frequency properties, while allowing individual segments to be placed within the substrate, adhering to crosstalk constraints. Furthermore, this partitioning allows us to navigate the complexity of chip real estate, particularly in densely populated areas where conventional rectangular resonator layouts are impractical.
Fig.\ref{fig:pad_parti}-d showcases potential reshaped patterns for resonators, \textit{It is important to clarify that the segmented blocks serve solely as placeholders to reserve space for the eventual placement of the resonators, rather than to physically partition the resonators themselves. The resonators are then re-routed based on the new positions of these segmented blocks.}
Fig.\ref{fig:pad_parti}-e illustrated the examples of actual resonator routing in Qiskit Metal \cite{Qiskit_Metal}, following the sequence established in Fig.\ref{fig:pad_parti}-d, where the red shadow indicates the space reserved. In the left figure of Fig.\ref{fig:pad_parti}-e, the target resonator navigates around an obstacle (highlighted in a red dotted box), demonstrating the adaptability and versatility of this method.
\textit{It is crucial to understand that resonator partitioning influences only the placement of the location of resonator, without altering its dimensional attributes such as gap width, etch depth, or metal heights. Moreover, our routing strictly adheres to the configurations in Qiskit Metal}\mbox{~\cite{Qiskit_Metal}} \textit{to avoid sharp corners that could lead to potential surface losses, thereby minimizing associated risks}\mbox{\cite{surface_loss, surface_loss_1, surface_loss_2}}.


\subsection{Quantum Analytical Placement}

The complexity of determining quantum component locations while adhering to crosstalk constraints presents significant computational challenges. Its inherent computational intricacy means even rudimentary tasks deemed NP-complete. To navigate this complexity, our methodology bifurcates the process into two distinct stages: placement and legalization.

\subsubsection{Frequency-aware Electrostatic-based Placement Engine}
Our proposed frequency-aware electrostatic-based placement engine is a pivotal component in achieving efficient substrate utilization and mitigating crosstalk in designing quantum device layout. This engine translates the constraints of crosstalk into an objective function, which is then iteratively minimized to approximate an optimal layout. Within this framework, both qubits and resonator segments are treated as movable instances $i, i\in I$, each characterized by unique frequency properties $\omega_i$

\begin{figure}[t]
  \centering
  \includegraphics[width=0.9\linewidth]{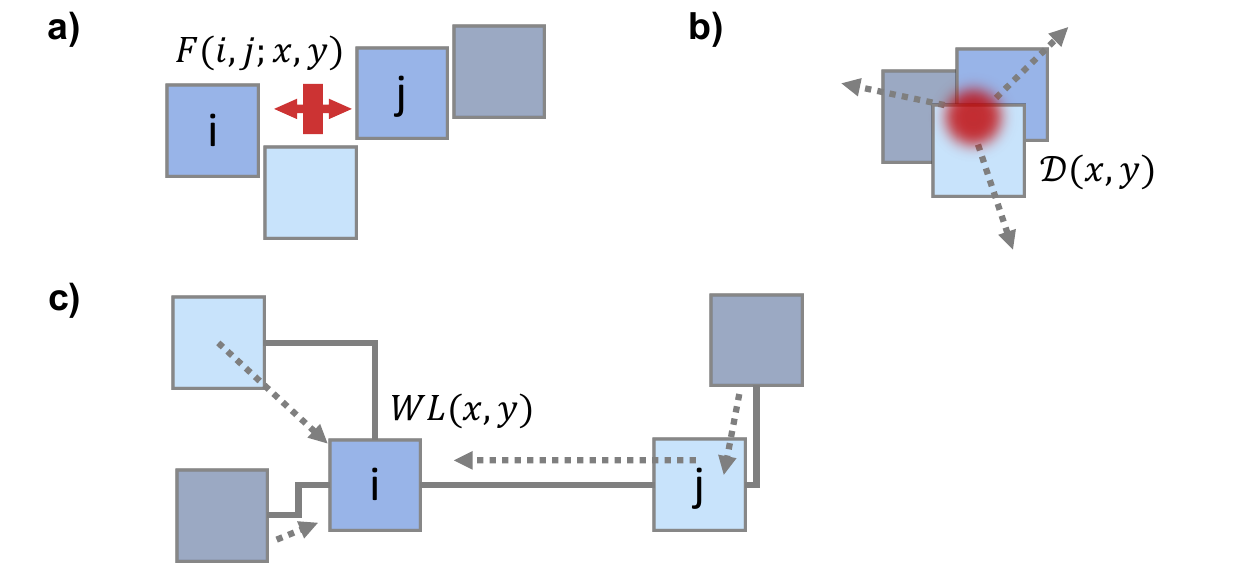}
  \caption{
  Frequency-aware placement engine visualization with instances color-coded by frequency:
  \textbf{a):} Frequency constraint $F$ repels resonant instances in close proximity (red arrows depict frequency repulsive forces).
  \textbf{b):} Density constraint $\mathcal{D}$ disperses instances to achieve target density $\hat{\mathcal{D}}$ (red dots represent areas of high density).
  \textbf{c):} Wirelength $WL$ optimization minimizes substrate area by drawing instances closer together.
  }
  \label{fig:method_placer}
  \vspace{-10pt}
\end{figure}

\textbf{Frequency Constraint:}
In addressing the spatial crosstalk constraints detailed in Section \ref{sec:problem}, quantum instances such as qubits and resonators necessitate to repel other instances that are on or near resonance in close distance proximity to prevent unintentional interactions. This phenomenon is analogous to charged entities repelling each other in an electrostatic field. Drawing inspiration from electrostatic systems, our model treats movable instances as charged entities, exerting a `\textbf{frequency repulse force}’ $F$ on each other when they share similar frequencies, as illustrated in Fig.\ref{fig:method_placer}-a.
\begin{equation}\label{eq:force}
    F(i, j; x, y) = \frac{\tau(\omega_i, \omega_j, \Delta_c)}{(x_i-x_j)^2+(y_i-y_j)^2}
\end{equation}
In Eq.(\ref{eq:force}), $x_n$ and $y_n$ denote instance $n$'s coordinates, while $\Delta_c$ represents the detuning threshold to avoid resonance. Resonance in subsequent discussions implies a detuning smaller than $\Delta_c$. The crosstalk indicator $\tau(\cdot)$ activates (1) for frequency differences within this threshold ($|\omega_i - \omega_j | \leq \Delta_c$) and deactivates (0) otherwise. This setup selectively applies force to instances with near-resonance frequencies, optimizing their positioning to reduce crosstalk.

To ensure that resonator segments, which are partitioned for layout flexibility, are not inappropriately dispersed, we modify our frequency repulse force as shown in Eq.(\ref{eq:force_constrain})
\begin{equation}\label{eq:force_constrain}
     F(i, j; x, y) = \frac{\tau(\omega_i, \omega_j, \Delta_c)}{(x_i-x_j)^2+(y_i-y_j)^2} \cdot (1-\delta(r_i, r_j))
\end{equation}
Where $r$ is the resonator index (only checking for resonator segment blocks). $\delta(\cdot)$ is the Kronecker delta function, which is 1 for identical frequencies and 0 otherwise. $(1-\delta(r_i, r_j))$ indicator ignores the instances from the same resonator. 

Note that, to efficiently compute the frequency repulse force $F$, a frequency collision map is constructed for each quantum instance before placement. This map identifies potential crosstalk instances, excluding those belonging to the same resonator. Typically, the map contains dozens to hundreds of instances for each quantum instance under consideration. During the placement optimization process, only the collision map is iterated through, avoiding the computationally expensive `all-to-all' iterations.

\textbf{Density Constraint:}
Frequency constrain only separate the instance with resonant frequency rather than impact the detuned instances, which results the illegally overlapped scenario among detuned instances. Density constrain is next introduced to resolve this problem, which restrict the density at any location on the substrate remains below a predefined target value $\hat{\mathcal{D}}$. 
\begin{equation}
    \mathcal{D}(x, y) \leq \hat{\mathcal{D}}
\end{equation}
Where $\mathcal{D}(x, y)$ represents the density at a given location in the layout. This constraint is interpreted as potential energy within the electrostatic analogy. The electric potential and field distribution are governed by Poisson's equation, incorporating charge density distribution as input \cite{eplace}.

\textbf{Objective Function and Optimization:} The ultimate objective of placement engine is to efficiently allocate all movable instances within the legal space of the quantum chip. To efficiently quantize the area usage, we utilize the wirelength concept, a common metric in classical placement, denoted as $\text{WL}(e, x, y)$. Here, $e$ in $E$ is a connection between two components, and this function measures the length of each wire $e$. This approach underpins the formulation of our comprehensive objective function.
\begin{equation}
    \min_{x,y} \sum_{i, j \in I; e \in E} \text{WL}(e; x, y),
\end{equation}

\begin{equation}
    \text{s.t.} \quad \mathcal{D}(x, y) \leq \hat{\mathcal{D}},\; F(i, j; x, y)
\end{equation}
Penalty method is adopted to efficiently solve this constrained optimization problem, by transforming it into a series of unconstrained problems. The transformed objective is:
\begin{equation}\label{gp_obj}
\min_{x,y} \sum_{i, j \in I; e \in E} \text{WL}(e;x,y) + \lambda \mathcal{D}(x,y) + \lambda_f F(i,j; x,y)
\end{equation}
$F$ and $\mathcal{D}$ denotes the frequency penalty and density penalty, respectively, which facilitates the proper dispersion of instances across the layout. To meet these constraints, the parameter $\lambda$ and $\lambda_f$ are increased progressively. With smaller parameters, the placement engine focus is predominantly on reducing the area regarding wirelength. As these parameters escalates, there is a seamless shift from prioritizing area minimization to achieving a balance between area and constrain penalty optimization. Consequently, movable instances disperse methodically, ensuring the area remains optimal.

\begin{figure}[t]
  \centering
  \includegraphics[width=0.9\linewidth]{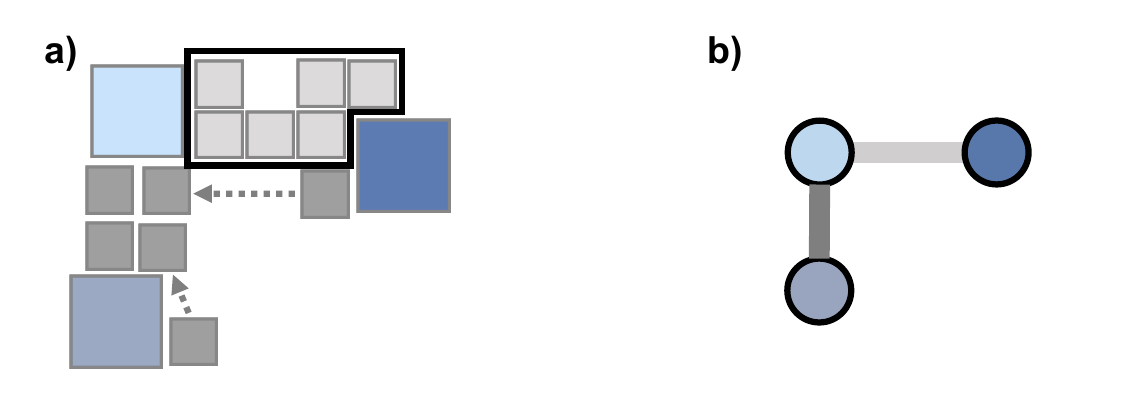}
  \caption{
  \textbf{a):} Resonator integration legalizer fixes legalized resonator and integrate the remaining resonator from its largest segment cluster. connectivity topology is shown in \textbf{b)}
  }
  \label{fig:method_lg}
  \vspace{-10pt}
\end{figure}

\subsubsection{Integration-Aware Legalization}
The final phase of \name{} entails an integration-aware legalization, designed to legalize component arrangement. 
The process commences with qubit legalization, temporarily sidelining segmented resonator blocks. 
This involves a greedy spiral search to identify viable, non-overlapping locations for movable qubits \cite{dreamplace}, followed by a min-cost flow refinement to minimize displacement \cite{macro}. 
Once qubit positions are set, attention shifts to arranging segmented resonator blocks using a modified Tetris-like methodology \cite{tetris}, placing blocks from left to right ensuring minimal displacement and adherence to established orders. 

\begin{algorithm}[t]
    \caption{Integration-Aware Legalization} \label{alg:lg}
    \begin{algorithmic}[1]
    \small
    \Require Placement solution $(x_i, y_i)$ for all instance $I$, $\forall i \in I$, the segment list $v_r$ for each resonator $R$, $\forall r \in R$, All qubits $Q$, $R\bigcup Q = I$, resonant checker $\tau(\cdot)$, qubit legalizer $\text{Q-LG}(\cdot)$, segment Tetris legalizer $\text{T-LG}(\cdot)$, and resonator integration legality checker $\text{rilc}(\cdot)$
    
    \Ensure legalized placement solution with resonator integration
    
    \State $x_q, y_q \gets \text{Q-LG}(x_q, y_q), \forall q \in Q 
    \quad\quad\quad \triangleright$ Legalize qubits
    \State $x_i, y_i \gets \text{T-LG}(x_i, y_i), \forall i \in I, i \notin Q \quad\triangleright$ Legalize segments
    \For{$v_r \in R$}
        \If{$rilc(v_r)$}
            \State $x_i, y_i \gets \forall i \in v_r \quad\triangleright$ fix segments
        \Else
            \State $C_{v_r} \gets v_r \quad\triangleright$ find largest segment cluster 
        \EndIf
    \EndFor
    \State $I_u \gets I, R_u \gets R \quad\triangleright$ get unfixed segments and resonators
    \For{$v_r \in R$}
        \State $C_{v_r}\cdot\tau(C_{v_r}) \gets C_{v_r}, I_u \quad\triangleright$ swap segments
        \If{$rilc(C_{v_r})$}
            \State $x_i, y_i \gets \forall i \in v_r \quad\triangleright$ fix segments
        \EndIf
    \EndFor
    
    \end{algorithmic}
\end{algorithm}

Post-Tetris refinement, it involves a thorough evaluation of the placement, particularly concentrating on the integration of resonator segments. The criterion for successful integration mandates that all segments of a resonator should be in close proximity to at least one other segment from the same resonator. 
Each resonator is examined to verify its compliance with this criterion. For resonators that comply, their segments position are finalized. Conversely, for those that fall short, we find its largest contiguous segment clusters and initiate a swapping process between the neighboring instances of cluster and scattered segments associated resonant checking $\tau(\cdot)$, aiming to enlarger the clusters. 
This procedure ensures that resonators can be seamlessly reassembled from its constituent segments. The detailed legalization flow is described in Algorithm \ref{alg:lg}.

\section{Evaluation Methodology}
\subsection{Benchmarks}
We conduct performance evaluations using a range of quantum device connectivity topologies and NISQ benchmarks, as outlined in Table \ref{tab:benchmark}. These topologies, which are either prevalent in industrial applications or designed for algorithmic efficiency, vary in qubit count, ranging from 25 to 127. This diversity in quantum device structures allows us to thoroughly assess our engine's scalability and adaptability across different quantum processor configurations.

\subsection{Evaluation Comparisons}
The following placement schemes are assessed in several comparative evaluations to test the performance of our frequency-aware electrostatic-based quantum placer. 

\begin{itemize}[leftmargin=*]
    \item \textbf{Human}: 
    The optimal manually optimized, crosstalk-free design represented by IBM’s layout, positions qubits in a 2D grid according to the connectivity topology, minimizing total edge length. Qubits are connected only to their nearest neighbors, with resonators strategically placed to prevent intersections or overlaps. The distance $D$ between two qubits is determined by the resonator’s area, which is calculated as the product of its length ($L$) and padding size ($d_r$). The resonator’s area is then reshaped to align with the padded qubit size ($L_q + 2d_q$), yielding $D = \frac{L \times d_r}{L_q + 2d_q}$ \cite{ibm_127_3, quantware}.

    \item \textbf{Classic}: The state-of-the-art Classical placement engine \cite{dreamplace} uses default hyperparameter settings and incorporates resonator partitioning technique detailed in Section \ref{sec:partition}.
    
    \item \textbf{Qplacer}: Our proposed frequency-aware electrostatic-based quantum placement engine. For fair comparison, it adheres to the same hyperparameter settings as \textbf{Classic} baseline.

\end{itemize}

\begin{table}[t] 
  \centering
  \footnotesize
  \caption{TOPOLOGIES AND BENCHMARKS}

  \begin{tabular}{lcl}
    \toprule
    Topology & Qubits & Description \\
    \midrule
    Grid      & 25 & \scriptsize{Quantum error correction friendly architecture\cite{google}}\\
    Heavy Hex & 27  & \scriptsize{Falcon processor from IBM \cite{ibmq}}\\
    Heavy Hex & 127 & \scriptsize{Eagle processor from IBM \cite{ibmq}}\\
    Octagon   & 40  & \scriptsize{Aspen-11 processor from Rigetti \cite{rigetti}}\\
    Octagon   & 80  & \scriptsize{Aspen-M processor from Rigetti \cite{rigetti}}\\ 
    Xtree     & 53  & \scriptsize{Pauli-String efficient architecture in Level 3 \cite{xtree}}\\

    \midrule
    \midrule
    
    Benchmark & Qubits & Description \\
    \midrule
    BV    & 4, 9, 16 & \scriptsize{Bernstein-Varzirani algorithm \cite{bv}}\\
    QAOA  & 4, 9     & \scriptsize{Quantum Approximate Optimization Algorithm \cite{QAOA}} \\
    Ising & 4        & \scriptsize{Linear Ising model simulation of spin chain \cite{gate_time2}}\\
    QGAN  & 4, 9     & \scriptsize{Quantum Generative Adversarial Network \cite{QGAN}}\\

  \bottomrule
  
    \end{tabular}
\label{tab:benchmark}
\vspace{-10pt}
\end{table}


\subsection{Experiment setup}
\textbf{Tool Implementation:}
The framework was developed using Python, utilizing PyTorch for optimizers and APIs, and C++ for low-level operations. This development was based on the open-source placement engine \cite{dreamplace}. All experimental data were obtained from a Linux machine equipped with an Intel(R) Xeon(R) CPU E5-2687W v4 @ 3.00GHz.

\textbf{Architectural Features:} We utilize standard symmetric frequency-fixed pocket transmon qubits as shown in Fig.\ref{fig:transmon}-a, each having $400\times400 \mu m^2$ referred from Qiskit Metal\cite{Qiskit_Metal} with nearly constant aharmonicity $\alpha/2\pi = (\omega_{12}-\omega_{01})/2\pi \approx 310$MHz \cite{ibmq}. The available qubit frequency range $\Omega$ is set to realistic values 4.8$\sim$5.2 Ghz in line with experimental data from devices \cite{ibmq}.
The available resonator frequency range $\Omega_r$ is set to realistic values 6.0$\sim$7.0 Ghz in line with experimental data from literature \cite{cpw_spur, rip_gate, rip_gate_2, bus_resonator_setup_2}.
The resonator lengths correspond to frequencies within 10.8$\sim$9.2 mm, calculated using $f = v_0/2L$, where $v_0 \sim 1.3 \times 10^8 m/s$ is the speed of light in wavelength \cite{cQED}.

To comprehensive evaluate impact of spatial isolation with catering realistic variation in fabrication, the padding length for qubit and resonator are aggressively set to $d_q=400 \mu m$ and $d_r=100\mu m$, respectively (the minimum distance between two adjacent components are the sum of their paddings).
While detuning threshold $\Delta_c$ is set to $0.1$GHz. These setups result reasonable coupling strength for any detuned qubits, but higher coupling strength between two resonance qubits, as illustrated in Fig.\ref{fig:g_q_dist} and \ref{fig:g_res_res}-c. 
We adopt a 3D packaging approach, focusing on qubits and bus resonators while omitting readout resonators \mbox{\cite{ibm_127_1, ibm_127_2}}. Ground plane is currently not considered.

\textbf{Metrics:} 
To assess crosstalk susceptibility in our placement experiments, we analyse the layout quality from three perspectives: general algorithm program fidelity, area and frequency hotspots proportion. 

\noindent\textbf{(1) General algorithm program fidelity:} program fidelity $\mathcal{F}$ is modeled using three terms to estimate the worst case fidelity of a program benchmark under crosstalk and decoherence noises, similar to the fidelity metric used in \cite{crosstalk_ding, google}:
\begin{align}
    \mathcal{F} = 
    \Pi_{q \in Q} (1-\epsilon_q) \cdot 
    \Pi_{g \in G}(1-\epsilon_g) \cdot 
    \Pi_{r \in R}(1-\epsilon_{r})
    \label{eq:fidelity}
\end{align}
where $\epsilon_q$ represents the qubit error from single qubit gates, two-qubit gates and decoherence, decoherence error represent by decay constants $T_1$ and $T_2$ during both idle and gate operation periods \cite{coherence_time_3}.
The crosstalk gate error for qubits, $\epsilon_g$, occurs when two qubits violate spatial constraints, akin to being linked by a direct capacitive coupling as detailed in Section \ref{sec:q_x}. 
This error results from Rabi oscillations, periodic energy exchanges between the qubits (state $\ket{01}$ and $\ket{10}$, state $\ket{11}$ and $\ket{20}$), driven by their effective coupling strength $g_{\text{eff}}$. The transition probability is modeled as $\Pr[t] = \sin^2(g_{\text{eff}}t)$ \cite{gate_time1, crosstalk_ding}, the corresponding crosstalk error for idle qubits is:
\begin{equation}
    \epsilon_g(\Delta,t) = 1-\sin{(g_{\text{eff}}(\Delta)t)}^2 
    \label{eq:eg}
\end{equation}
Similarly, $\epsilon_r$ accounts for crosstalk errors among resonators under spatial violations, analogous to those measured for qubits. The parasitic capacitance depends on the adjacent length.
It is crucial to note that these fidelity calculations pertain only to actively engaged physical qubits and resonators in layouts, as errors in inactive elements do not compromise the program’s overall fidelity.

\noindent\textbf{(2) Area:} We define the layout's area using two parameters. $A_{\text{mer}}(I)$, which is the area of the minimum enclosing rectangle that encompasses all instances, providing a measure of the layout's overall spatial requirements; $A_{\text{poly}}(I)$ represents the total area covered by all individual instances, calculated as the sum of their polygonal areas. The substrate area utilization ratio is determined by:
\begin{equation}
    \frac{A_{\text{poly}}(I)}{A_{\text{mer}}(I)}
\end{equation}
which quantifies the efficiency of space usage within the defined layout perimeter.

\noindent\textbf{(3) Frequency Hotspot Proportion:} we introduce a novel metric, \textit{Frequency Hotspot Proportion ($P_h$)} to quantify the potential for crosstalk by identifying `\textbf{frequency hotspots}'. These are defined as regions where instances are closely positioned with frequency differences smaller than a predefined threshold $\Delta_c$, which signifies a heightened risk of crosstalk. The metric is calculated as follows:
\begin{equation}
    P_{h} = \frac{\sum_{i, j \in I} (p_i \cap p_j) \cdot d_c(p_i, p_j) \cdot \tau(\omega_i, \omega_j, \Delta_c)}{A_{poly}(I)}
\end{equation}
Here, $p_n$ represents the polygon of instance $n$, $p_i \cap p_j$ is the intersection length between two instances, $d_c(p_i, p_j)$ denotes the centroid distance between their polygons, and $\tau$ is a function that assesses frequency proximity based on the threshold $\Delta_c$. This metric effectively translates crosstalk impacts into measurable spatial terms, facilitating a detailed analysis of crosstalk risks in quantum computing layouts and helping identify vulnerable qubits. For simplicity, the term ``frequency hotspot" is often shortened to ``hotspot" in discussions.

\section{Results}
\begin{figure*}[ht!]
    \centering
    \includegraphics[width=\linewidth]{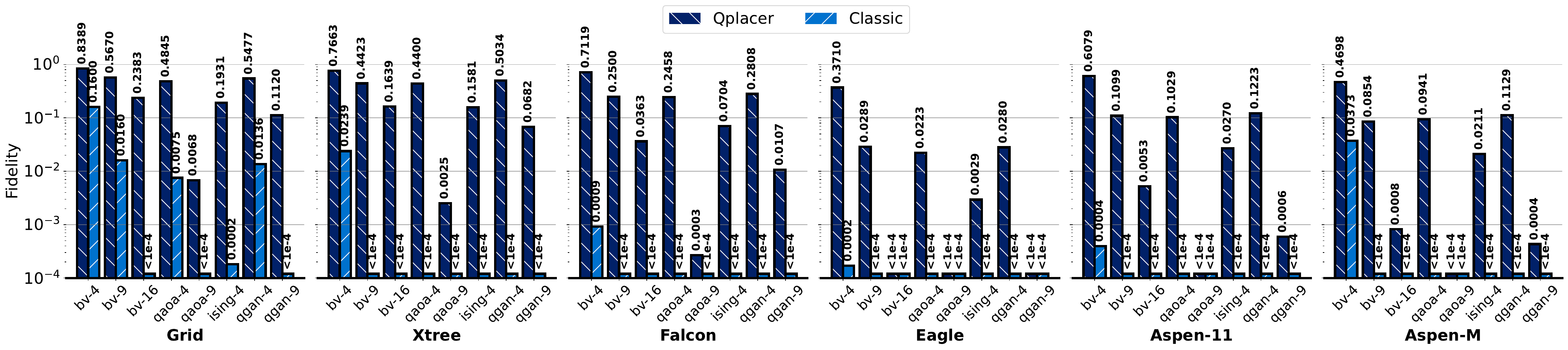}
    \vspace{-15pt}
    \caption{
    Fidelity estimation from various placement strategies. A higher fidelity value indicates a better performance. Across all benchmarks, \name{} consistently outperforms the Classic baseline, demonstrating its superior efficacy in maintaining higher fidelity levels in layouts.}

    \label{fig:exp_fidelity}
    \vspace{-15pt}
\end{figure*}
\begin{figure}
    \centering
    \includegraphics[width=0.95\linewidth]{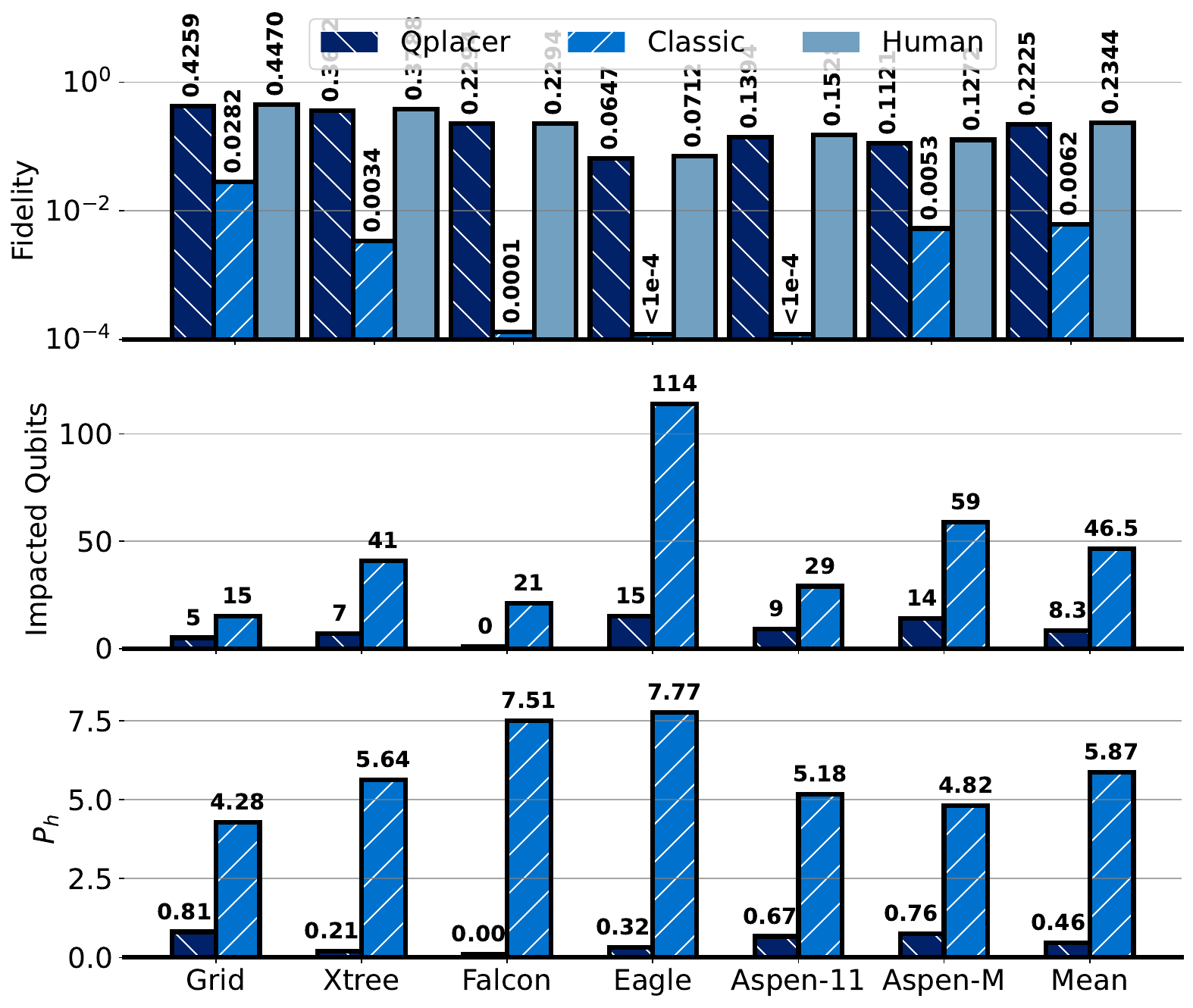}
    \vspace{-5pt}
    \caption{
    A comparison of different placement schemes based on three metrics: average program fidelity, number of impacted qubits, and the proportion of frequency hotspots. For the latter two metrics, lower values indicate better performance. Across all tested topologies, Qplacer consistently excels in reducing frequency hotspots.
    }
    \label{fig:exp_hotspots}
    \vspace{-15pt}
\end{figure}

\subsection{Fidelity Estimation}\label{sec:fidelity}

Fig.\ref{fig:exp_fidelity} presents the worst-case overall fidelity for various placement strategies, all subjected to the same hyper-parameters and estimated using our noise model—Eq.(\ref{eq:fidelity}). 
And the upper subfigure in 
Fig.\ref{fig:exp_hotspots} is the average value of Fig.\ref{fig:exp_fidelity} across the topologies.
The size of the resonator segments, $l_b$, is fixed at 0.3 mm, which was found to be optimal as detailed in Section \ref{sec:sweep_lb}. 
Given that the number of required qubits for the application is fewer than the physical qubits available on each topology, each mapping evaluation is confined to a subset of the device’s qubits.
To rigorously evaluate each layout, we conducted evaluations on 50 different subsets of physical qubits, aiming to encompass all the physical qubits across different topological layouts. We utilized Qiskit’s mapping with L3 optimization to enhance the mappings, drawing on basis gates from IBMQ \cite{ibmq} to minimize the impacts from software-level by reducing circuit depth and gate numbers.
This approach allows for the assessment of performance consistency across all qubits within a chip, addressing concerns that a few mapping samples might not accurately reflect the overall system performance.For consistency and fair comparison, the same mappings were used across all benchmarks and placers. Each bar in the figure represents the average fidelity of these 50 mappings for a benchmark on a specific topology layout.

Classic placer, which does not account for crosstalk, generally suffers from a significant reduction in fidelity across most circuits and topologies, managing only to produce moderate layouts for simpler topologies with fewer qubits such as grid. In contrast, \name consistently achieves superior performance, improving fidelity by an average of 36.7\x{} across all benchmarks and topologies, as illustrated in the upper subfigure of Fig.\ref{fig:exp_hotspots}.
This includes complex benchmarks like QAOA-9, where \name still maintains relatively high fidelity across various topologies. The success of \name over classic strategies can be attributed to its heuristic approach of optimizing instance positions based on frequency relationships.

\subsection{\FH Proportion}\label{sec:FH}
Fig.\ref{fig:exp_hotspots} primarily illustrates the proportion of frequency hotspots, $P_h$, and the number of qubits affected by these hotspots under different placement strategies. \name{}, which employs a frequency-aware placement engine and an integration legalizer, excels in achieving spatial isolation for movable instances, with an average violation rate of only 0.46\%. In contrast, the classic placement engine, which does not consider crosstalk, exhibits a higher average hotspot proportion of 5.87\%, indicating that \name{} is approximately 12.76\x{} more effective in reducing crosstalk-related spatial violation.

The correlation between program fidelity and $P_{h}$ is depicted in the subfigures of Fig.\ref{fig:exp_hotspots}, with the upper subfigure showing average benchmark fidelities for each topology and the lower subfigure displaying $P_{h}$. These visuals indicate that program fidelity is inversely proportional to $P_{h}$, validating the effectiveness of $P_{h}$ as a metric for assessing layout quality.

Further analysis reveals that the number of qubits impacted by hotspots is not strictly linear increase but tends to be exponential with the hotspot proportion, highlighting the non-localized nature of resonator crosstalk; even a minor misplacement can significantly affect the fidelity of all connected qubits. For instance, in the Eagle layout using a classical placement strategy, while the hotspot proportion is about 7.77\%, this condition adversely affects over 110 qubits, which constitutes more than 90\% of the total qubits (127). Such findings underscore the critical importance of strategic placement in mitigating the pervasive impact of crosstalk.

\subsection{Area Optimization} \label{sec:area}
\begin{figure}
    \centering
    \includegraphics[width=0.95\linewidth]{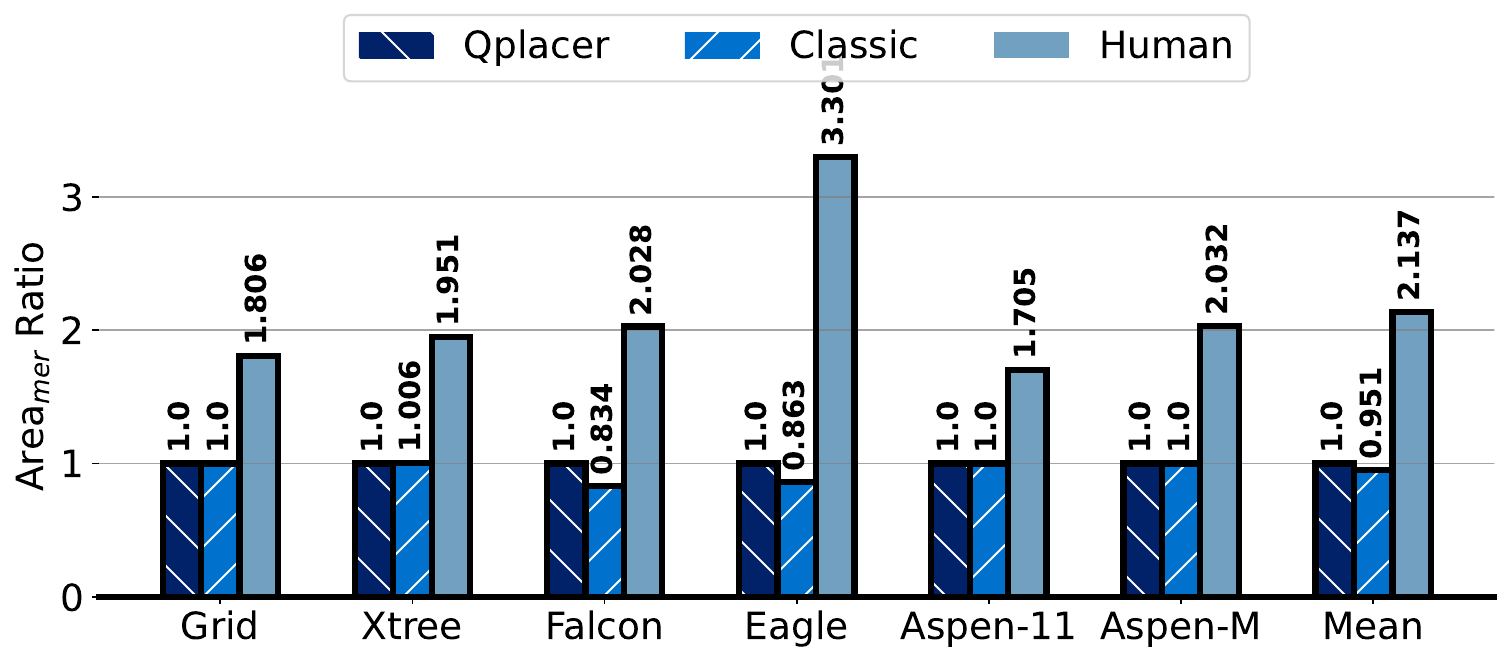}
    \vspace{-5pt}
    \caption{
   Minimum enclosing rectangle area ($A_{\text{mer}}$) ratios of placement schemes relative to \name{}. A smaller ratio is preferred as it indicates a more compact layout. \name provides consistent benefits, outperforming the competing schemes. 
    }
    \label{fig:exp_area}
    \vspace{-15pt}
\end{figure}
\begin{figure}[t]
    \centering
    \includegraphics[width=0.9\linewidth]{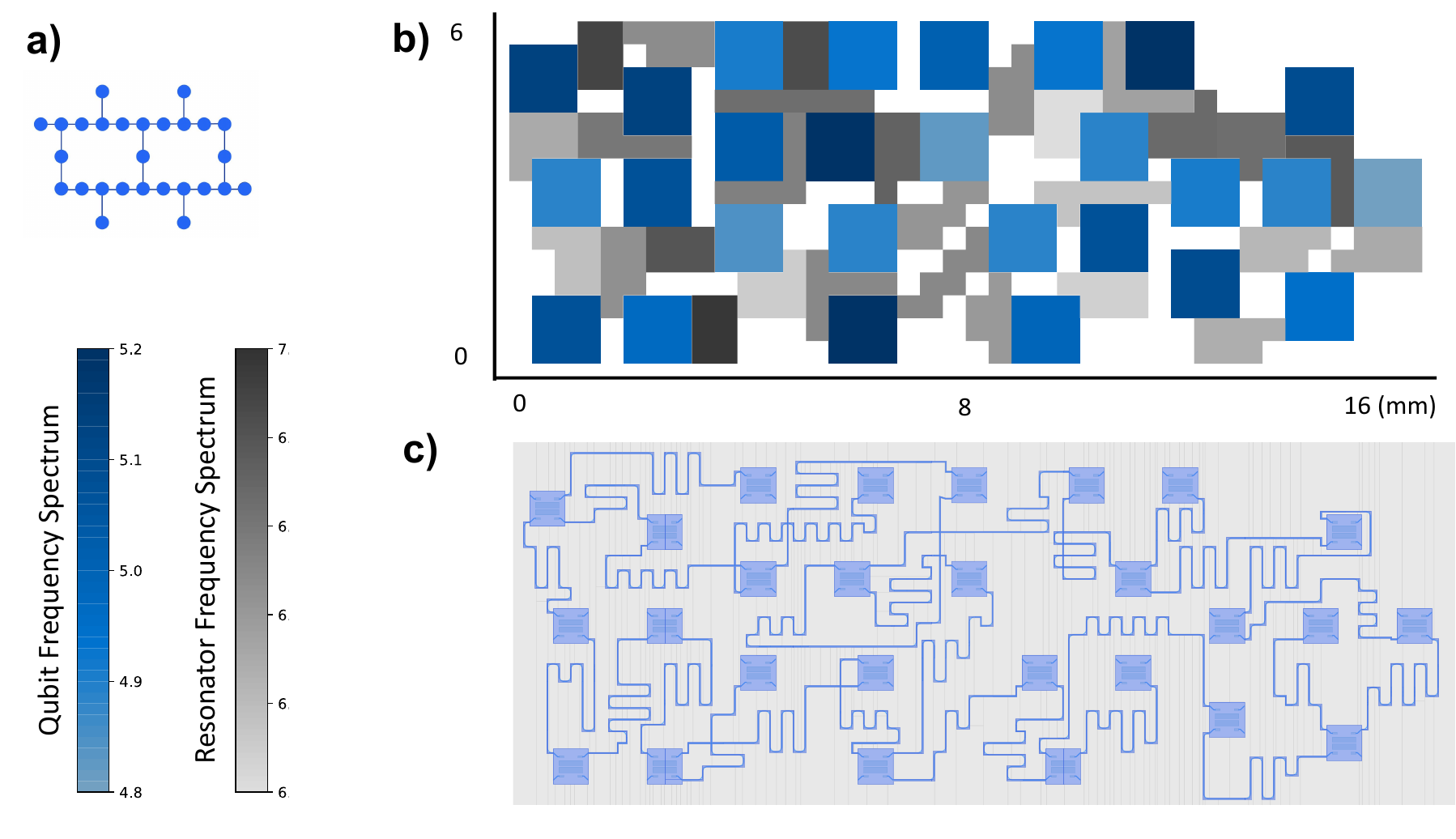}
    \vspace{-10pt}
    \caption{
    \textbf{a):} Input frequency spectrums and connectivity topology (Falcon);
    \textbf{b):} Optimized layout by \name.
    \textbf{c):} Generated GDS file from Qiskit Metal \cite{Qiskit_Metal} based on the optimized layout in \textbf{(b)}.
    }
    \label{fig:eagle}
    \vspace{-15pt}
\end{figure}

Fig.\ref{fig:exp_area} offers a comparative analysis of the minimum substrate area required ($A_{\text{mer}}$) across various placement strategies, benchmarked against Qplacer's $A_{\text{mer}}$. Human-designed layouts, which typically feature qubits arranged in a 2D grid with considerable spacing (as seen in systems such as those by \cite{quantware, ibmq}), often occupy larger substrate areas. Although these layouts are effective in minimizing crosstalk among on-chip instances, they may increase the likelihood of substrate spurious effects, potentially degrading system fidelity.

Layouts generated by classical placement engines have areas comparable to those produced by \name{}, since they share the same hyper-parameters. In stark contrast, \name{} enhances substrate area utilization, achieving on average a 2.14\x improvement compared to handcrafted layouts. This efficiency underscores \name{}'s dual capability in optimizing layout dimensions and enhancing fidelity. By balancing frequency constraints with wirelength optimization, \name{} consistently delivers compact layouts with reduced hotspot occurrences.

Fig.\ref{fig:eagle}-b presents a layout prototype for Falcon \cite{ibmq}. Despite some residual unoccupied space, \name{} efficiently minimizes the required area while maintaining fidelity. The gray areas denote space reserved for resonators, as detailed in Section \ref{sec:partition}. 
This prototype serves as a reference for users to finetune the layout and effectively route the resonators.
Moreover, an example layout in GDSII format, derived from Fig.\ref{fig:eagle}-b using Qiskit Metal, is displayed in Fig.\ref{fig:eagle}-c to illustrate the practical application of our optimized layout prototype.

\subsection{Sweeping the Segment Size}\label{sec:sweep_lb}
\begin{figure}[t]
    \centering
    \includegraphics[width=0.95\linewidth]{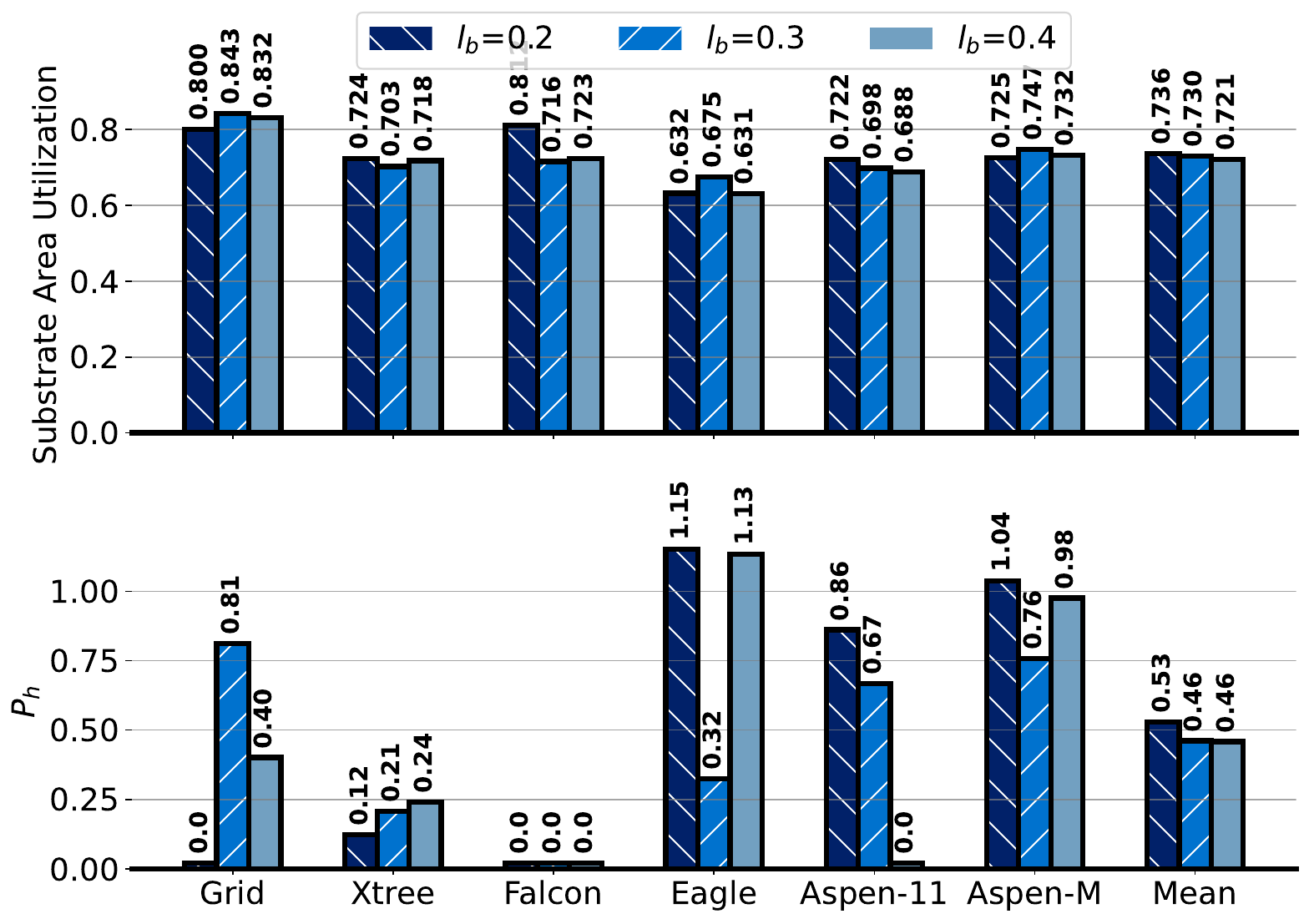}
    \vspace{-5pt}
    \caption{
   Substrate area utilization and hotspot proportion $P_h$ for \name with different resonator segment sizes $l_b$ (in mm).
    }
    \label{fig:exp_ab_blksize}
    \vspace{-10pt}
\end{figure}

This section evaluates the performance of resonator partitioning by exploring the impact of various resonator segment sizes ($l_b$) on layout optimization and placement runtime. The outcomes of the layout evaluations are depicted in Fig.\ref{fig:exp_ab_blksize}, while the corresponding placement runtimes (measured in seconds), average runtimes (in seconds), and the counts of instances (\#cell) are detailed in Table \ref{tab:exe_time}.

We analyzed three segment sizes: $l_b=0.2\text{, }0.3$, and $0.4$ mm. Analysis identified $l_b=0.3$ mm as the optimal segment size for these topologies, as it offers the best balance between substrate utilization and hotspot proportions, trading less than 1\% in area utilization for a 16.3\% reduction in hotspots. Furthermore, as shown in the last row of Table \ref{tab:exe_time}, using $l_b=0.2$ mm increases the number of instances by factors of 2.1\x{} and 3.5\x{} compared to $l_b=0.3$ mm and $l_b=0.4$ mm, respectively. This increase in instances correlates with runtime increases of 2.7\x{} and 3.87\x{}, respectively. These results highlight that the smallest segment size ($l_b=0.2$ mm) may not always be the most efficient choice; while it enhances area efficiency, it also leads to a higher number of instances, longer runtimes, and increased hotspot proportions.

Runtime data in Table \ref{tab:exe_time} also demonstrates scalability of \name{} in handling an increased number of qubits efficiently, enabling users to generate optimized layouts within seconds.

\begin{table}[t!] 
  \centering
  \caption{
  COMPARISON ON PLACEMENT TIME, \textit{RT} represents runtime in seconds, \textit{Avg} represents the average runtime per iteration
  }
  \scriptsize
  \setlength{\tabcolsep}{4pt} 
  \begin{tabular}{c|ccc|ccc|ccc}
    \toprule
    Topology & \multicolumn{3}{c|}{$l_b=0.2$} & \multicolumn{3}{c}{$l_b=0.3$} & \multicolumn{3}{|c}{$l_b=0.4$} \\
     & \#cells & RT & Avg & \#cells & RT & Avg & \#cells & RT & Avg \\
    \midrule
    Grid   & 1050 & 8.1  & 0.024 &  490 & 4.6  & 0.017 &  299 & 4.6 & 0.017 \\
    Xtree  & 1393 & 15.0 & 0.035 &  660 & 5.2  & 0.017 &  410 & 4.9 & 0.017 \\
    Falcon & 744  & 7.6  & 0.020 &  354 & 6.4  & 0.017 &  218 & 3.7 & 0.017 \\
    Eagle  & 3810 & 56.1 & 0.165 & 1801 & 11.3 & 0.044 & 1104 & 7.0 & 0.024 \\
    \tiny{Aspen-11} & 1272 & 12.4 & 0.030 &  598 & 7.2  & 0.018 &  369 & 6.5 & 0.017 \\
    \tiny{Aspen-M}	& 2787 & 28.5 & 0.093 & 1310 & 12.5 & 0.030 &  799 & 6.4 & 0.019 \\
    Mean   & 1843 & 21.3 & 0.061 & 869 & 7.9 & 0.023 & 533 & 5.5 &  0.018\\
  \bottomrule
\end{tabular}
\label{tab:exe_time}
\vspace{-15pt}
\end{table} 

\section{RELATED WORK}
Crosstalk in quantum computing has been a critical issue addressed by numerous studies, with primarily focusing on inter-qubit interactions. Strategies to mitigate this involve the use of compilers and schedulers in fixed coupling systems \cite{ibmq, fix_xtalk_miti, VQP, majority} and the incorporation of tunable components in more adaptive architectures \cite{crosstalk_ding, gate_time1, tunable_qubit, tunable_qubit_2, google}. These approaches largely center on temporal or frequency isolation to prevent crosstalk \cite{crosstalk, crosstalk_ding, ibm_crosstalk, disq, csar}. However, the importance of spatial isolation is often underemphasized, which leads to detrimental parasitic couplings \cite{para_g, para_g_2, para_g_3}.

In terms of resonator-based crosstalk, modifications in hardware design, such as material selection and frequency detuning, have been explored \cite{resonator_crosstalk, substrate_limit, res_crosstalk}. Another significant factor is substrate-induced crosstalk, which escalates with increased substrate size, leading to spurious electromagnetic modes \cite{spurious, dist_modeling, cpw_spur}. Efforts to enhance substrate scalability include advanced packaging techniques \cite{substrate_limit_2, substrate_limit, packaging_1}, through-silicon vias (TSV) integration \cite{dist_modeling, tsv_1, tsv_2, tsv_3}, high-dielectric substrates \cite{packaging_1, substrate_design_1, dielectric_loss}, air-bridge cross-overs \cite{airbridges}, and flip-chip lids \cite{flip_chip}. Despite these advancements, spurious modes remain a bottleneck, limiting substrate size and thus impacting quantum computing scalability \cite{scale, cQED, quantum_progress_2}. 
The concept of a quantum chiplet model \cite{chiplet, ibm_dqs} has been introduced as a potential solution to these scalability issues.

From the perspective of substrate utilization for scaling up quantum processors, placement engines play crucial roles. 
Electrostatic-based global placement algorithms are widely used due to their effectiveness in providing smooth density penalty functions and comprehensive views of placement zones at granular levels \cite{replace, eplace, eplace_ms, dreamplace, dreamplace3}. Following placement, legalization processes such as greedy search \cite{macro}, Tetris-like approaches \cite{tetris}, and row-based Abacus refinement \cite{abacus} are critical in addressing overlaps.

\section{CONCLUSION}
We introduced \name{}, a frequency-aware, electrostatic-based placement framework tailored for robust and scalable superconducting quantum processors. Our approach strategically positions quantum components on substrates, optimizing the usage of limited area while preserving system fidelity, particularly mitigating diverse crosstalk impacts. 

Notably, other techniques like compilation and transpilation methods, can be applied orthogonally to \name{} to further enhance system robustness.
While our primary focus was on fixed-frequency architectures, its versatile design renders it suitable for a wide array of quantum architectures, including those with tunable elements which often share similar geometrical configurations with fixed ones.
In addition, this work highlights the placement problem within the quantum computing community, a critical but previously overlooked issue as quantum systems scale. 
placement strategy not only resolves crosstalk (Our focus here) but also has the potential to mitigate other challenges like decoherence by incorporating additional chip components like airbridges, vias.

\newpage


\bibliographystyle{IEEEtranS}
\bibliography{references}

\end{document}